\documentclass[journal,a4paper,oneside,onecolumn]{IEEEtran} 
\usepackage{amsfonts,amssymb}
\usepackage[utf8]{inputenc} 
\usepackage[T1]{fontenc}
\usepackage{url}
\usepackage{ifthen}
\usepackage{cite}
\usepackage[cmex10]{amsmath}
\usepackage{mathabx}
\newtheorem{thm}{Theorem}
\newtheorem{lem}{Lemma}
\newtheorem{df}{Definition}
\newtheorem{rem}{Remark}
\newtheorem{example}{Example}

\usepackage{curve2e}

\usepackage{algorithm}
\usepackage{algpseudocode}
\makeatletter
\let\OldStatex\Statex
\renewcommand{\Statex}[1][3]{%
 \setlength\@tempdima{\algorithmicindent}%
 \OldStatex\hskip\dimexpr#1\@tempdima\relax}
\makeatother
\algnewcommand{\IIf}[1]{\State\algorithmicif\ #1\ \algorithmicthen}
\algnewcommand{\FFor}[1]{\State\algorithmicfor\ #1\ \algorithmicdo}
\algnewcommand{\EndFFor}{\unskip\ \algorithmicend\ \algorithmicfor}

\newcommand{\lrB}[1]{\left[{#1}\right]}
\newcommand{\lrb}[1]{\left\{{#1}\right\}}
\newcommand{\lrsb}[1]{\left({#1}\right)}
\newcommand{\lrbar}[1]{\left|{#1}\right|}

\newcommand{\A}{\mathcal{A}}
\newcommand{\C}{\mathcal{C}}
\newcommand{\D}{\mathcal{D}}
\newcommand{\E}{\mathcal{E}}
\newcommand{\F}{\mathcal{F}}
\newcommand{\G}{\mathcal{G}}
\newcommand{\I}{\mathcal{I}}
\newcommand{\J}{\mathcal{J}}
\newcommand{\K}{\mathcal{K}}
\newcommand{\M}{\mathcal{M}}
\newcommand{\R}{\mathcal{R}}
\newcommand{\cS}{\mathcal{S}}
\newcommand{\T}{\mathcal{T}}
\newcommand{\U}{\mathcal{U}}
\newcommand{\V}{\mathcal{V}}
\newcommand{\W}{\mathcal{W}}
\newcommand{\X}{\mathcal{X}}
\newcommand{\Y}{\mathcal{Y}}
\newcommand{\Z}{\mathcal{Z}}

\newcommand{\fC}{\mathfrak{C}}
\newcommand{\fI}{\mathfrak{I}}
\newcommand{\fL}{\mathfrak{L}}

\newcommand{\NN}{\mathbb{N}}

\newcommand{\CC}{\boldsymbol{C}}
\newcommand{\MM}{\boldsymbol{M}}
\newcommand{\UU}{\boldsymbol{U}}
\newcommand{\VV}{\boldsymbol{V}}
\newcommand{\WW}{\boldsymbol{W}}
\newcommand{\XX}{\boldsymbol{X}}
\newcommand{\YY}{\boldsymbol{Y}}
\newcommand{\ZZ}{\boldsymbol{Z}}

\newcommand{\cc}{\boldsymbol{c}}
\newcommand{\mm}{\boldsymbol{m}}
\newcommand{\bp}{\boldsymbol{p}}
\newcommand{\uu}{\boldsymbol{u}}
\newcommand{\vv}{\boldsymbol{v}}
\newcommand{\xx}{\boldsymbol{x}}
\newcommand{\yy}{\boldsymbol{y}}
\newcommand{\zz}{\boldsymbol{z}}

\newcommand{\bcF}{\boldsymbol{\F}}

\newcommand{\zero}{\boldsymbol{0}}
\newcommand{\one}{\boldsymbol{1}}
\newcommand{\aalpha}{\boldsymbol{\alpha}}
\newcommand{\bbeta}{\boldsymbol{\beta}}

\newcommand{\hu}{\widehat{u}}
\newcommand{\hM}{\widehat{M}}
\newcommand{\hU}{\widehat{U}}
\newcommand{\hZ}{\widehat{Z}}
\newcommand{\hmm}{\widehat{\mm}}
\newcommand{\hzz}{\widehat{\zz}}
\newcommand{\hZZ}{\widehat{\ZZ}}

\newcommand{\tZ}{\widetilde{Z}}

\newcommand{\chu}{\check{u}}
\newcommand{\chU}{\check{U}}
\newcommand{\chZ}{\check{Z}}
\newcommand{\chzz}{\check{\zz}}

\newcommand{\ocS}{\stackrel{\circ}{\cS}}
\newcommand{\dcS}{\stackrel{\bullet}{\cS}}

\newcommand{\oO}{\overline{O}}
\newcommand{\oQ}{\overline{Q}}
\newcommand{\oH}{\overline{H}}
\newcommand{\uH}{\underline{H}}
\newcommand{\oT}{\overline{\mathcal{T}}}
\newcommand{\uT}{\underline{\mathcal{T}}}

\newcommand{\e}{\varepsilon}

\newcommand{\limn}{\lim_{n\to\infty}}
\newcommand{\liminfn}{\liminf_{n\to\infty}}
\newcommand{\pliminfn}{\operatornamewithlimits{\text{p-liminf}}_{n\to\infty}}
\newcommand{\plimsupn}{\operatornamewithlimits{\text{p-limsup}}_{n\to\infty}}

\newcommand{\im}{\mathrm{Im}}
\newcommand{\Prod}{\operatornamewithlimits{\text{\Large $\times$}}}

\newcommand{\markov}{\leftrightarrow}
\newcommand{\Prob}{\mathrm{P}}
\newcommand{\Error}{\mathrm{Error}}

\newcommand{\RIT}{\R_{\mathrm{IT}}}
\newcommand{\ROP}{\R_{\mathrm{OP}}}

\allowdisplaybreaks

\title{
 Channel Codes for Relayless Networks\\
 with General Message Access Structure
}
\author{
 \IEEEauthorblockN{Jun~Muramatsu~\IEEEmembership{Senior Member,~IEEE}}
 \\
  \IEEEauthorblockA{
   NTT Communication Science Laboratories, NTT Corporation\\
   2-4 Hikaridai, Seika-cho, Soraku-gun, Kyoto, 619-0237 Japan\\
   E-mail: jun.muramatsu@ieee.org.
  }
}
\begin{document}
\maketitle

\begin{abstract}
 Channel codes for relayless networks
 with the general message access structure
 is introduced.
 It is shown that
 the multi-letter characterized capacity region
 of this network is achievable with this code.
 The capacity region is characterized in terms of entropy functions
 and provides an alternative to the regions introduced by
 [Somekh-Baruch and Verd\'u, ISIT2006][Muramatsu and Miyake, ISITA2018].
\end{abstract}

\section{Introduction}

\begin{figure}[ht]
\begin{center}
 \unitlength 0.52mm
 \begin{picture}(157,85)(-5,10)
  \put(-6,90){\makebox(0,0){$M^{(n)}_1$}}
  \put(-6,66){\makebox(0,0){$M^{(n)}_2$}}
  \put(-6,38){\makebox(0,0){$\vdots$}}
  \put(-6,6){\makebox(0,0){$M^{(n)}_{|\cS|}$}}
  \put(0,90){\vector(20,-11){20}}
  \put(0,90){\vector(20,-70){20}}
  \put(0,66){\vector(20,12){20}}
  \put(0,66){\vector(20,-14){20}}
  \put(0,6){\vector(20,14){20}}
  \put(0,6){\vector(20,71){20}}
  \put(27,92){\makebox(0,0){Encoders}}
  \put(20,68){\framebox(14,20){$\Phi^{(n)}_1$}}
  \put(20,42){\framebox(14,20){$\Phi^{(n)}_2$}}
  \put(27,38){\makebox(0,0){$\vdots$}}
  \put(20,10){\framebox(14,20){$\Phi^{(n)}_{|\I|}$}}
  \put(34,78){\vector(1,0){10}}
  \put(49,78){\makebox(0,0){$X^n_1$}}
  \put(54,78){\vector(1,0){10}}
  \put(34,52){\vector(1,0){10}}
  \put(49,52){\makebox(0,0){$X^n_2$}}
  \put(54,52){\vector(1,0){10}}
  \put(49,38){\makebox(0,0){$\vdots$}}
  \put(34,20){\vector(1,0){10}}
  \put(49,20){\makebox(0,0){$X^n_{|\I|}$}}
  \put(54,20){\vector(1,0){10}}
  \put(64,10){\framebox(24,78){\small $\mu_{Y^n_{\J}|X^n_{\I}}$}}
  \put(127,92){\makebox(0,0){Decoders}}
  \put(88,78){\vector(1,0){10}}
  \put(104,78){\makebox(0,0){$Y^n_1$}}
  \put(110,78){\vector(1,0){10}}
  \put(120,68){\framebox(14,20){$\Psi^{(n)}_1$}}
  \put(88,52){\vector(1,0){10}}
  \put(104,52){\makebox(0,0){$Y^n_2$}}
  \put(110,52){\vector(1,0){10}}
  \put(120,42){\framebox(14,20){$\Psi^{(n)}_2$}}
  \put(127,38){\makebox(0,0){$\vdots$}}
  \put(88,20){\vector(1,0){10}}
  \put(104,20){\makebox(0,0){$Y^n_{|\J|}$}}
  \put(110,20){\vector(1,0){10}}
  \put(120,10){\framebox(14,20){$\Psi^{(n)}_{|\J|}$}}
  \put(134,78){\vector(1,0){10}}
  \put(153,78){\makebox(0,0){$\hM^{(n)}_{\D(1)}$}}
  \put(134,52){\vector(1,0){10}}
  \put(153,52){\makebox(0,0){$\hM^{(n)}_{\D(2)}$}}
  \put(153,38){\makebox(0,0){$\vdots$}}
  \put(134,20){\vector(1,0){10}}
  \put(153,20){\makebox(0,0){$\hM^{(n)}_{\D(|\J|)}$}}
 \end{picture}
\end{center}
\caption{Multi-terminal Channel Coding}
\label{fig:channel}
\end{figure}
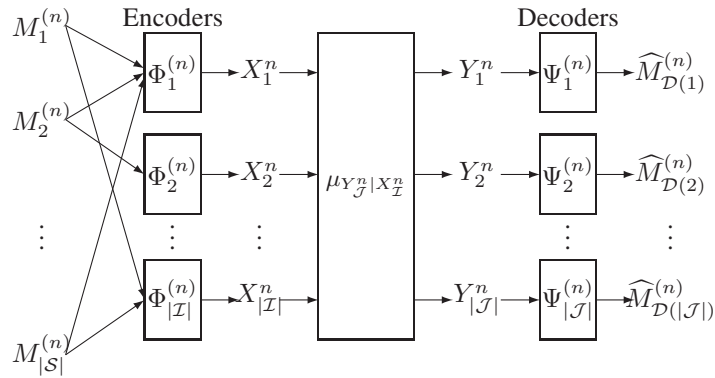

This paper investigates the problem of multi-terminal channel coding
for relayless networks with the general message access structure
shown in Fig.~\ref{fig:channel}.
Multi-terminal channels include
broadcast channels~\cite{C72}\cite{GP80}\cite{IO05}\cite{LKP11}\cite{M79}\cite{ITW13},
multiple-access channels~\cite{A71}\cite{A74}\cite{EC80}\cite{H79}\cite{H98}\cite{L72}\cite{SHANNON61}\cite{SW73MAC},
and interference channels~\cite{A74}\cite{CMGE08}\cite{HK81}\cite{JXG08}\cite{ICC}.

The contribution of this paper is the introduction
of codes for this type of network
by using constrained-random-number generators,
which are the basic building blocks for the construction
of the both encoders and decoders.
Sparse matrices (with logarithmic column degree) are available
for code construction.
The construction includes
the case when all messages are private \cite{CRNG-MULTI}
and the case when all encoders have access to all common messages \cite{ICC}.
It should be noted that
there is an unsupported case in which previous constructions \cite{ICC}\cite{CRNG-MULTI} cannot be applied directly.

It is shown that the multi-letter characterized capacity region
of this network is achievable with this code.
This capacity region is specified in terms of entropy functions
and provides an alternative to the region derived
in~\cite{CRNG-MULTI}\cite{SV06}.
It should be noted that,
when random variables are assumed to be stationary and memoryless,
our region provides the best known single-letter characterized
achievable regions
for general stationary memoryless channels,
where the rate-splitting technique is unnecessary~\cite{ITW13}\cite{ICC}.

Throughout this paper, we use the following definitions and notations.
When $\U$ is a set and $\V_u$ is also a set for each $u\in\U$,
we use the notation $\V_{\U}\equiv\Prod_{u\in\U}\V_u$.
We use the notation $v_{\U}\equiv\{v_u\}_{u\in\U}\in\V_{\U}$
to represent the sequence of elements
(e.g.\ sequences, random variables, functions)
$v_u$ with index $u\in\U$.
We use the notation $|\U|$ to represent the cardinality of $\U$.
Let $2^{\U}$ be the power set of $\U$.

Let $\I$ be the index set of channel inputs,
and $\J$ be the index set of channel outputs.
Then, a general channel is characterized 
by sequence $\{\mu_{Y^n_{\J}|X^n_{\I}}\}_{n=1}^{\infty}$
of conditional distributions,
where $n\in\NN$ is the block length of the channel input,
$X^n_{\I}\equiv\{X^n_i\}_{i\in\I}$
is the set of random variables of multiple channel inputs,
and $Y^n_{\J}\equiv\{Y^n_j\}_{j\in\J}$
is the set of random variables of multiple channel outputs.

For each $i\in\I$ and $n\in\NN$,
let $\X_i^n$ be the alphabet of random variable
$X_i^n\equiv(X_{i,1},\ldots,X_{i,n})$,
where we assume that
$\X_i^n$ is the $n$-dimensional Cartesian product of finite set $\X_i$
and $X_{i,k}\in\X_i$ for all $k\in\{1,\ldots,n\}$.
For each $j\in\J$ and $n\in\NN$,
let $\Y_j^n$ be the alphabet of random variable $Y_j^n$.
It should be noted that
$\Y_j^n$ is allowed to be an infinite/continuous set
and it is unnecessary to assume that
$\Y_j^n$ is the $n$-dimensional Cartesian product of $\Y_j$.
For example, we can assume that $\Y^n_j\equiv\bigcup_{n=0}^{\infty}\X_i^n$,
which is the set of all finite length sequences with alphabet $\X_i$,
to describe insertion-deletion-substitution channels.
We use notations $Y_j^n$ and $\Y_j^n$
to consider the stationary memoryless case.

Let $\cS$ be the index set of multiple messages.
For each $s\in\cS$ and $n\in\NN$,
let $M^{(n)}_s$ be a random variable of Message $s$
corresponding to the uniform distribution on alphabet $\M^{(n)}_s$.
We assume that $\{M^{(n)}_s\}_{s\in\cS}$ are mutually independent.
We consider the situation that
each encoder has access to some of the messages in $\{M^{(n)}_s\}_{s\in\cS}$,
where some messages are common to some encoders.
The definition of the general message access structure
between messages and encoders is introduced in Section~\ref{sec:access}.

\section{Message Access Structure}
\label{sec:access}

This section introduces the message access structure.

{\em (Message) access structure} $\A$ is a subset of $\cS\times\I$,
where member $(s,i)\in\A$ indicates that
Encoder $i$ has access to Message~$s$.
It should be noted that $(\cS,\I,\A)$ forms a directed bipartite graph,
where $(s,i)\in\A$ corresponds to the arc (directed edge) $s\to i$.
For each $s\in\cS$, let $\I(s)$ be the set of all indices of encoders
that have access to Message $s$, where $\I(s)$ is defined as
\begin{equation*}
 \I(s)\equiv\{i\in\I: (s,i)\in\A\}.
\end{equation*}
For each $i\in\I$, let $\cS(i)$ be the set of all indices of the messages
to which Encoder $i$ has access, where $\cS(i)$ is defined as
\begin{equation*}
 \cS(i)\equiv\{s\in\cS: (s,i)\in\A\}.
\end{equation*}
We have the fact that $i\in\I(s)$ is equivalent to $s\in\cS(i)$.

For a given $\I'\in2^{\I}$,
we refer to the set of encoders whose index belongs to $\I'$
as {\em Encoders $\I'$}.
Let $\cS(\I')$ be the index set of messages common to Encoders $\I'$,
where $\cS(\I')$ is defined as
\begin{equation}
 \cS(\I')
 \equiv
 \{
  s\in\cS: \I(s)=\I'
  \}.
 \label{eq:SI}
\end{equation}
We define $\fI$ as
\begin{equation*}
 \fI
 \equiv
 \{\I'\in 2^{\I}: \cS(\I')\neq\emptyset\}.
\end{equation*}

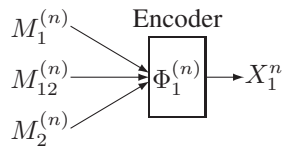
\begin{figure}[ht]
\begin{center}
 \unitlength 0.52mm
 \begin{picture}(57,53)(-5,10)
  \put(0,46){\makebox(0,0)[r]{$M^{(n)}_1$}}
  \put(0,33){\makebox(0,0)[r]{$M^{(n)}_{12}$}}
  \put(0,20){\makebox(0,0)[r]{$M^{(n)}_2$}}
  \put(0,46){\vector(20,-12){20}}
  \put(0,33){\vector(20,0){20}}
  \put(0,20){\vector(20,12){20}}
  \put(27,48){\makebox(0,0){Encoder}}
  \put(20,23){\framebox(14,20){$\Phi^{(n)}_1$}}
  \put(34,33){\vector(1,0){10}}
  \put(49,33){\makebox(0,0){$X^n_1$}}
 \end{picture}
\end{center}
\caption{Access Structure of Example \ref{ex:1}}
\label{fig:example1}
\end{figure}

Here, let us introduce a few examples.
\begin{example}[Broadcast channel with a common message]
 \label{ex:1}
 The access structure
 of a broadcast channel with a common message (Fig.~\ref{fig:example1})
 can be written as
 \begin{align*}
  \cS&\equiv\{1,2,12\}
  \\
  \I&\equiv\{1\}
  \\
  \A&\equiv\{(1,1),(2,1),(12,1)\},
 \end{align*}
 where
 Encoder $1$ has access to Messages $1$, $2$, and $12$,
 Message $12$ is reproduced by Decoders $1$ and $2$,
 and Message $i$ is reproduced by Decoder $i$ for each $i\in\{1,2\}$.
 We have
 \begin{align*}
  \cS(1)&\equiv\{1,2,12\}
  \\
  \I(1)&\equiv\{1\}
  \\
  \I(2)&\equiv\{1\}
  \\
  \I(12)&\equiv\{1\}
  \\
  \fI&\equiv\{\{1\}\}
  \\
  \cS(\{1\})&\equiv\{1,2,12\}.
 \end{align*}
\end{example}

\begin{figure}[ht]
\begin{center}
 \unitlength 0.52mm
 \begin{picture}(57,65)(-5,10)
  \put(0,46){\makebox(0,0)[r]{$M^{(n)}_1$}}
  \put(0,33){\makebox(0,0)[r]{$M^{(n)}_{12}$}}
  \put(0,20){\makebox(0,0)[r]{$M^{(n)}_2$}}
  \put(0,46){\vector(1,0){20}}
  \put(0,33){\vector(20,12){20}}
  \put(0,33){\vector(20,-12){20}}
  \put(0,20){\vector(1,0){20}}
  \put(27,60){\makebox(0,0){Encoders}}
  \put(20,36){\framebox(14,20){$\Phi^{(n)}_1$}}
  \put(20,10){\framebox(14,20){$\Phi^{(n)}_2$}}
  \put(34,46){\vector(1,0){10}}
  \put(49,46){\makebox(0,0){$X^n_1$}}
  \put(34,20){\vector(1,0){10}}
  \put(49,20){\makebox(0,0){$X^n_2$}}
 \end{picture}
\end{center}
\caption{Access Structure of Example \ref{ex:2}}
\label{fig:example2}
\end{figure}
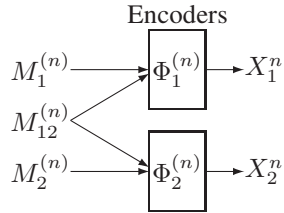

\begin{example}[Two-input multiple access channel with a common message]
 \label{ex:2}
 The access structure of a two-input multiple-access channel
 with a common message (Fig.~\ref{fig:example2})
 can be written as
 \begin{align*}
  \cS&\equiv\{1,2,12\}
  \\
  \I&\equiv\{1,2\}
  \\
  \A&\equiv\{(1,1),(2,2),(12,1),(12,2)\},
 \end{align*}
 where Message $12$ is a common message for Encoders $1$ and $2$,
 and Message $i$ is a private message for Encoder $i$ for each $i\in\{1,2\}$.
 In other words, Encoder $i$ has access to Messages $i$ and $12$
 for each $i\in\{1,2\}$.
 This access structure is the same as that of
 the two-user interference channel with a common message \cite{JXG08},
 where Decoder $i$ reproduces Messages $i$ and $12$ for each $i\in\{1,2\}$.
 We have
 \begin{align*}
  \cS(1)&\equiv\{1,12\}
  \\
  \cS(2)&\equiv\{2,12\}
  \\
  \I(1)&\equiv\{1\}
  \\
  \I(2)&\equiv\{2\}
  \\
  \I(12)&\equiv\{1,2\}
  \\
  \fI
  &\equiv
  \{
   \{1,2\},
   \{1\},
   \{2\}
   \}
  \\
  \cS(\{1,2\})
  &\equiv
  \{12\}
  \\
  \cS(\{1\})
  &\equiv
  \{1\}
  \\
  \cS(\{2\})
  &\equiv
  \{2\}.
 \end{align*}
\end{example}

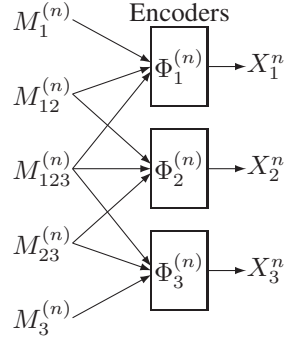
\begin{figure}[ht]
\begin{center}
 \unitlength 0.52mm
 \begin{picture}(57,85)(-5,10)
  \put(0,84){\makebox(0,0)[r]{$M^{(n)}_1$}}
  \put(0,65){\makebox(0,0)[r]{$M^{(n)}_{12}$}}
  \put(0,46){\makebox(0,0)[r]{$M^{(n)}_{123}$}}
  \put(0,27){\makebox(0,0)[r]{$M^{(n)}_{23}$}}
  \put(0,8){\makebox(0,0)[r]{$M^{(n)}_3$}}
  \put(0,84){\vector(20,-11){20}}
  \put(0,65){\vector(20,7){20}}
  \put(0,46){\vector(20,25){20}}
  \put(0,65){\vector(20,-18){20}}
  \put(0,46){\vector(1,0){20}}
  \put(0,27){\vector(20,18){20}}
  \put(0,46){\vector(20,-25){20}}
  \put(0,27){\vector(20,-7){20}}
  \put(0,8){\vector(20,11){20}}
  \put(27,86){\makebox(0,0){Encoders}}
  \put(20,62){\framebox(14,20){$\Phi^{(n)}_1$}}
  \put(20,36){\framebox(14,20){$\Phi^{(n)}_2$}}
  \put(20,10){\framebox(14,20){$\Phi^{(n)}_3$}}
  \put(34,72){\vector(1,0){10}}
  \put(49,72){\makebox(0,0){$X^n_1$}}
  \put(34,46){\vector(1,0){10}}
  \put(49,46){\makebox(0,0){$X^n_2$}}
  \put(34,20){\vector(1,0){10}}
  \put(49,20){\makebox(0,0){$X^n_3$}}
 \end{picture}
\end{center}
\caption{Access Structure of Example \ref{ex:3}}
\label{fig:example3}
\end{figure}

\begin{example}
 \label{ex:3}
 Here, we introduce an access structure
 of a multiple-access channel with three inputs
 (Fig.~\ref{fig:example3});
 it is written as
 \begin{align*}
  \cS&\equiv\{1,3,12,23,123\}
  \\
  \I&\equiv\{1,2,3\}
  \\
  \A&\equiv\lrb{
   \begin{aligned}
    &(1,1),(3,3),
    \\
    &(12,1),(12,2),(23,2),(23,3),
    \\
    &(123,1),(123,2),(123,3)
   \end{aligned}
  },
 \end{align*}
 where Message $i$ is a private message for Encoder $i$
 for each $i\in\{1,3\}$,
 Message $ij$ is a common message to Encoders $i$ and $j$
 for each two-digit indexes $ij\in\{12,23\}$,
 and Message $123$ is a common message to Encoders $1$, $2$, and $3$.
 In other words,
 Encoder $1$ has access to Messages $1$, $12$, and $123$,
 Encoder $2$ has access to Messages $12$, $23$, and $123$,
 and Encoder $3$ has access to Messages $3$, $23$, and $123$.
 It should be noted that
 there are partially-common messages $12$ and $23$,
 that do not appear in two-input multiple access channels.
 We have
 \begin{align*}
  \cS(1)
  &\equiv\{1,12,123\}
  \\
  \cS(2)
  &\equiv\{12,23,123\}
  \\
  \cS(3)
  &\equiv\{3,23,123\}
  \\
  \I(1)
  &\equiv\{1\}
  \\
  \I(3)
  &\equiv\{3\}
  \\
  \I(12)
  &\equiv\{1,2\}
  \\
  \I(23)
  &\equiv\{2,3\}
  \\
  \I(123)
  &\equiv\{1,2,3\}
  \\
  \fI
  &\equiv
  \{
   \{1,2,3\},
   \{1,2\},
   \{2,3\},
   \{1\},
   \{3\}
   \}
  \\
  \cS(\{1,2,3\})
  &\equiv
  \{123\}
  \\
  \cS(\{1,2\})
  &\equiv
  \{12\}
  \\
  \cS(\{2,3\})
  &\equiv
  \{23\}
  \\
  \cS(\{1\})
  &\equiv
  \{1\}
  \\
  \cS(\{3\})
  &\equiv
  \{3\}.
 \end{align*}
 It should be noted that
 this example is an unsupported case
 in which previous constructions \cite{ICC}\cite{CRNG-MULTI}
 cannot be applied directly.
\end{example}

From the following two lemmas,
we have the fact that $\{\cS(\I')\}_{\I'\in\fI}$ forms a partition of $\cS$.
\begin{lem}
\label{lem:union}
\begin{equation*}
 \bigcup_{\I'\in\fI}\cS(\I')=\cS.
\end{equation*}
\end{lem}
\begin{IEEEproof}
 Since $\bigcup_{\I'\in\fI}\cS(\I')\subset\cS$ is trivial,
 it is sufficient to show that $\cS\subset\bigcup_{\I'\in\fI}\cS(\I')$.
 Assume that $s\in\cS$ and $\I'\equiv\I(s)$.
 Then we have $s\in\cS(\I(s))=\{s':\I(s')=\I(s)\}$.
 This implies that $\cS(\I(s))\neq\emptyset$ and $\I'=\I(s)\in\fI$.
 Then we have $s\in\bigcup_{\I'\in\fI}\cS(\I')$
 and $\cS\subset\bigcup_{\I'\in\fI}\cS(\I')$.
\end{IEEEproof}
\begin{lem}
\label{lem:disjoint}
For any $\I'$ and $\I''$ satisfying $\I'\neq\I''$,
we have
\begin{equation*}
 \cS(\I')\cap\cS(\I'')=\emptyset.
\end{equation*}
\end{lem}
\begin{IEEEproof}
 We show the lemma by contradiction.
 Assume that $\I'\neq\I''$ and $\cS(\I')\cap\cS(\I'')\neq\emptyset$.
 From $\cS(\I')\cap\cS(\I'')\neq\emptyset$,
 there is $s\in\cS(\I')\cap\cS(\I'')$ satisfying $\I(s)=\I'$ and $\I(s)=\I''$.
 Then we have $\I'=\I''$, which contradicts $\I'\neq\I''$.
\end{IEEEproof}

Let $\ocS(\I')$ and $\dcS(\I')$ be defined as
\begin{align}
 \ocS(\I')
 &\equiv
 \bigcup_{\substack{
   \I''\in\fI:
   \\
   \I''\supsetneq\I'
 }}
 \cS(\I'')
 \label{eq:oSI}
 \\
 \dcS(\I')
 &\equiv
 \bigcup_{\substack{
   \I''\in\fI:
   \\
   \I''\subset\I'
 }}
 \cS(\I'').
 \label{eq:dSI}
\end{align}
Then we have the following lemmas.
\begin{lem}
\label{lem:disjoint-SIk-oSIk}
For any $\I'\in\fI$, we have
\begin{equation*}
 \cS(\I')\cap\ocS(\I')=\emptyset.
\end{equation*}
\end{lem}
\begin{IEEEproof}
 The lemma is shown immediately from Lemma~\ref{lem:disjoint}.
\end{IEEEproof}
\begin{lem}
\label{lem:cap-Si}
For any $\I'\in\fI$, we have
\begin{equation*}
 \bigcap_{i\in\I'}\cS(i)
 =
 \bigcup_{\substack{
   \I''\in\fI:
   \\
   \I''\supset\I'
 }}
 \cS(\I'')
 =
 \ocS(\I')\cup\cS(\I').
\end{equation*}
\end{lem}
\begin{IEEEproof}
The relation 
$\bigcup_{\I''\in\fI:\I''\supset\I'}\cS(\I'')=\ocS(\I')\cup\cS(\I')$
is shown immediately from the definition of $\ocS(\I')$.
We show below the relations
$\bigcap_{i\in\I'}\cS(i)\subset\bigcup_{\I''\in\fI:\I''\supset\I'}\cS(\I'')$
and
$\bigcup_{\I''\in\fI:\I''\supset\I'}\cS(\I'')
\subset\bigcap_{i\in\I'}\cS(i)$; together they imply
$\bigcap_{i\in\I'}\cS(i)=\bigcup_{\I''\in\fI:\I''\supset\I'}\cS(\I'')$.

First, assume that $s\in\bigcap_{i\in\I'}\cS(i)$ and $\I''\equiv\I(s)$.
Then we have the fact that $s\in\cS(i)$ for all $i\in\I'$.
Since $s\in\cS(i)$ implies $i\in\I(s)$,
we have the fact that $\I'\subset\I(s)=\I''$.
Since $s\in\cS(\I(s))=\{s':\I(s')=\I(s)\}$,
we have $\cS(\I(s))\neq\emptyset$ and $\I''=\I(s)\in\fI$.
Then we have $s\in\bigcup_{\I''\in\fI:\I''\supset\I'}\cS(\I'')$,
which implies
$\bigcap_{i\in\I'}\cS(i)\subset\bigcup_{\I''\in\fI:\I''\supset\I'}\cS(\I'')$.

Next, assume that $s\in\bigcup_{\I''\in\fI:\I''\supset\I'}\cS(\I'')$.
Then there is $\I''$ such that $\I''\supset\I'$ and $s\in\cS(\I'')$.
Since $s\in\cS(\I'')$ implies $\I'\subset\I''=\I(s)$,
we have $i\in\I(s)$ and $s\in\cS(i)$ for all $i\in\I'$.
Then we have $s\in\bigcap_{i\in\I'}\cS(i)$,
which implies
$\bigcup_{\I''\in\fI:\I''\supset\I'}\cS(\I'')\subset\bigcap_{i\in\I'}\cS(i)$.
\end{IEEEproof}
\begin{lem}
 \label{lem:subsetSi}
 For any $i\in\I'$,
 we have
 \begin{align}
  \cS(\I')
  &\subset
  \cS(i)
  \label{eq:SIsupSi}
  \\
  \ocS(\I')
  &\subset
  \cS(i).
  \label{eq:SsupIsupSi}
 \end{align}
 The above relations imply that
 Encoder $i$ has access to the set of messages
 $M^{(n)}_{\cS(\I')}\equiv\{M^{(n)}_s\}_{s\in\cS(\I')}$
 and $M^{(n)}_{\ocS(\I')}\equiv\{M^{(n)}_s\}_{s\in\ocS(\I')}$.
\end{lem}
\begin{IEEEproof}
 The lemma is shown immediately from Lemma \ref{lem:cap-Si}.
\end{IEEEproof}

\begin{lem}
\label{lem:unionSIk=Si}
For any $i\in\I$, we have
\begin{equation*}
 \bigcup_{\substack{
   \I'\in\fI:
   \\
   i\in\I'
 }}
 \cS(\I')=\cS(i).
\end{equation*}
\end{lem}
\begin{IEEEproof}
 First, we show $\bigcup_{\I'\in\fI: i\in\I'}\cS(\I')\subset\cS(i)$.
 Let $s\in\bigcup_{\I'\in\fI: i\in\I'}\cS(\I')$.
 Then there is $\I'\in\fI$ such that $i\in\I'$ and $s\in\cS(\I')$.
 From $s\in\cS(\I')$, we have $\I(s)=\I'$.
 Since $i\in\I'$ implies $i\in\I(s)$, we have $s\in\cS(i)$.
 Next, we show $\cS(i)\subset\bigcup_{\I'\in\fI: i\in\I'}\cS(\I')$.
 Let $s\in\cS(i)$ and $\I'\equiv\I(s)$.
 Then we have $i\in\I(s)=\I'$.
 Since $s\in\cS(\I(s))=\{s': \I(s')=\I(s)\}$,
 we have $\cS(\I(s))\neq\emptyset$ and $\I'=\I(s)\in\fI$.
 Then we have $s\in\bigcup_{\I'\in\fI: i\in\I'}\cS(\I')$.
 From the above two facts, we have the lemma.
\end{IEEEproof}

In subsequent sections,
we assume that all elements in $\fI\equiv\{\I_1,\I_2,\ldots,\I_{|\fI|}\}$
are sorted in a linear extension of the reversed partial ordering,
which yields the following property:
$\I_k\subsetneq\I_{k'}$ implies $k'<k$
for all $k,k'\in\{1,2,\ldots,|\fI|\}$.
In Examples \ref{ex:1}--\ref{ex:3},
all elements of $\fI$ are sorted in this order.
An algorithm for computing the linear extension
is described in Appendix \ref{sec:tsort}.
From (\ref{eq:oSI}), we have
\begin{align}
 \ocS(\I_k)
 =
 \bigcup_{\substack{
   k'\in\{1,\ldots,k-1\}:
   \\
   \I_{k'}\supsetneq\I_k
 }}
 \cS(\I_{k'}).
 \label{eq:ocSIk}
\end{align}
\begin{lem}
If $k'<k$, then
\[
 \cS(\I_{k'})\subset\dcS(\I_k)^{\complement}.
\]
\label{lem:k'<k}
\end{lem}
\begin{IEEEproof}
Let us assume that
$\cS(\I_{k'})\cap\dcS(\I_k)
=\cS(\I_{k'})\cap\lrB{\bigcup_{\I''\in\fI:\I''\subset\I_k}\cS(\I'')}
\neq\emptyset$.
Since Lemma \ref{lem:disjoint} imply that
only one of $\cS(\I_{k'})=\cS(\I'')$
and $\cS(\I_{k'})\cap\cS(\I'')=\emptyset$
holds,
we have the fact that $\I_{k'}\subset\I_k$.
Then we have $k'\geq k$ from the
assumption on $\{\I_1,\ldots,\I_{|\fI|}\}$.
Hence we have the fact that $k'<k$ implies 
$\cS(\I_{k'})\cap\dcS(\I_k)=\emptyset$
and 
$\cS(\I_{k'})\subset\dcS(\I_k)^{\complement}$.
\end{IEEEproof}

\section{Capacity Region}

This section introduces
the definition of a multi-letter characterized capacity region
for a general multiple-input-multiple-output channel coding \cite{CRNG-MULTI}.
Let $\Prob(\cdot)$ denote the probability of an event.

For each $i\in\I$, 
Encoder $i$ generates channel input $X^n_i$
from the set of messages $M^{(n)}_{\cS(i)}\equiv\{M^{(n)}_s\}_{s\in\cS(i)}$.
Decoder $j$ receives channel output $Y^n_j$
and reproduces the set of messages
$\hM^{(n)}_{\D(j)}\equiv\{\hM^{(n)}_{j,s}\}_{s\in\D(j)}$,
where $\D(j)$ is the index set of messages reproduced by Decoder $j$,
and $\hM^{(n)}_{j,s}$ is the reproduction by Decoder $j$
corresponding to Message $s$.
Let $\hM^{(n)}_{\D(\J)}\equiv\{\hM^{(n)}_{j,s}\}_{j\in\J,s\in\D(j)}$.
Then the joint distribution of
$(M^{(n)}_{\cS},X_{\I}^n,Y_{\J}^n,\hM^{(n)}_{\D(\J)})$ is given as
\begin{align*}
 &
 \mu_{M^{(n)}_{\cS}X_{\I}^nY_{\J}^n\hM^{(n)}_{\D(\J)}}(
  \mm_{\cS},\xx_{\I},\yy_{\J},\hmm_{\D(\J)}
 )
 \notag
 \\*
 &
 =
 \lrB{\prod_{j\in\J}\mu_{\hM^{(n)}_{\D(j)}|Y_j^n}(\hmm_{\D(j)}|\yy_{j})}
 \mu_{Y_{\J}^n|X_{\I}^n}(\yy_{\J}|\xx_{\I})
 \lrB{\prod_{i\in\I}\mu_{X_i^n|M^{(n)}_{\cS(i)}}(\xx_i|\mm_{\cS(i)})}
 \lrB{\prod_{s\in\cS}\frac 1{|\M^{(n)}_s|}}.
\end{align*}
We expect that, with probability close to $1$, $\hM^{(n)}_{j,s}=M^{(n)}_s$
for all $j\in\J$ and $s\in\D(j)$ if $n$ is sufficiently large.
We call rate vector $\{R_s\}_{s\in\cS}$ {\em achievable}
if there is a (possibly stochastic) code
$\{(\{\Phi^{(n)}_i\}_{i\in\I},\{\Psi^{(n)}_j\}_{j\in\J})\}_{n=1}^{\infty}$
consisting of encoders $\Phi^{(n)}_i:\M^{(n)}_{\cS(i)}\to\X^n_i$
and decoders $\Psi^{(n)}_j:\Y^n_j\to\M^{(n)}_{\D(j)}$ such that
\begin{gather}
 \liminfn\frac{\log_2 |\M^{(n)}_s|}n\geq R_s
 \quad\text{for all}\ s\in\cS
 \label{eq:ROP-rate}
 \\
 \limn
 \Prob\lrsb{
  \hM^{(n)}_{j,s}\neq M^{(n)}_s\ \text{for some}\ j\in\J\ \text{and}\ s\in\D(j)
 }
 =0,
 \label{eq:ROP-error}
\end{gather}
where $X_i^n\equiv\Phi^{(n)}_i(M^{(n)}_{\cS(i)})$
and $\hM^{(n)}_{\D(j)}\equiv\Psi^{(n)}_j(Y_j^n)$.
Capacity region $\ROP$ is defined as
the closure of the set of all achievable rate vectors.

In the following, we use the information spectrum method
introduced in \cite{HAN}, and we do not assume conditions such as consistency,
stationarity, and ergodicity.
For sequence $\{\mu_{U_nV_n}\}_{n=1}^{\infty}$ of
joint probability distributions corresponding to
$(\UU,\VV)\equiv\{(U_n,V_n)\}_{n=1}^{\infty}$,
$\uH(\UU|\VV)$ denotes the spectral conditional inf-entropy rate
and $\oH(\UU|\VV)$ denotes the spectral conditional sup-entropy rate.
Formal definitions are given in  Appendix~\ref{sec:ispec}.

Let $Z^n_{\cS}\equiv\{Z^n_s\}_{s\in\cS}$
be the random variables subject to the distribution defined as
\begin{align}
 p_{Z^n_{\cS}}(\zz_{\cS})
 &=
 \prod_{\substack{
   \I'\in\fI
 }}
 p_{Z^n_{\cS(\I')}|Z^n_{\ocS(\I')}}
 (\zz_{\cS(\I')}|\zz_{\ocS(\I')}),
 \label{eq:jointZ}
\end{align}
where
\begin{align*}
 p_{Z^n_{\cS(\I')}|Z^n_{\ocS(\I')}}
 (\zz_{\cS(\I')}|\zz_{\ocS(\I')})
 &\equiv
 p_{Z^n_{\cS(\I')}}(\zz_{\cS(\I')})
 \quad\text{if}\ \ocS(\I')=\emptyset.
\end{align*}
The alphabet of $Z^n_s$ is denoted by $\Z^n_s$ for each $s\in\cS$.
It should be noted that it is unnecessary to assume
that $\Z_s^n$ is the $n$-dimensional Cartesian product of $\Z_s$.
We use notations $Z_s^n$ and $\Z_s^n$ to consider
the stationary memoryless case.

Let $\RIT$  be defined as the set of all $\{R_s\}_{s\in\cS}$
satisfying the condition that
there are a set of general sources $\ZZ_{\cS}\equiv\{\ZZ_s\}_{s\in\cS}$
and a set of numbers $\{r_s\}_{s\in\cS}$ such that
\begin{align}
 R_s
 &\geq 
 0
 \label{eq:RIT-R}
 \\
 \sum_{s\in\cS'}
 [R_s+r_s]
 &\leq
 \uH(\ZZ_{\cS'}|\ZZ_{\ocS(\I')})
 \label{eq:RIT-rR}
 \\
 \sum_{s\in\D'}r_s
 &\geq
 \oH(\ZZ_{\D'}|\YY_j,\ZZ_{\D(j)\setminus\D'})
 \label{eq:RIT-r}
\end{align}
for all $(\I',\cS',j,\D')$ satisfying
$\I'\in\fI$, $\emptyset\neq\cS'\subset\cS(\I')$,
$j\in\J$, $\emptyset\neq\D'\subset\D(j)$,
where the joint distribution of $(Z^n_{\cS},X^n_{\I},Y_{\J}^n)$ is given as
\begin{align}
 \mu_{Z_{\cS}^nX_{\I}^nY_{\J}^n}(\zz_{\cS},\xx_{\I},\yy_{\J})
 &=
 \mu_{Y_{\J}^n|X_{\I}^n}(\yy_{\J}|\xx_{\I})
 \lrB{\prod_{i\in\I}\mu_{X_i^n|Z^n_{\cS(i)}}(\xx_i|\zz_{\cS(i)})}
 \mu_{Z^n_{\cS}}(\zz_{\cS})
 \label{eq:jointXYZ}
\end{align}
by using $\mu_{Z^n_{\cS}}$ defined by (\ref{eq:jointZ}).
It should be noted that we can eliminate auxiliary variables
$\{r_s\}_{s\in\cS}$ by applying the Fourier-Motzkin method \cite[Appendix D]{EK11}.

We have the following theorem.
The proof of $\ROP\subset\RIT$ is given in Section \ref{sec:converse}.
For the proof of $\ROP\supset\RIT$,
we construct a code in Section \ref{sec:channel-code}.
\begin{thm}
\label{thm:ROP=RIT}
\begin{equation*} 
 \ROP=\RIT.
\end{equation*}
\end{thm}

\begin{rem}
When channels and auxiliary sources are stationary and memoryless,
we can replace $\uH(\ZZ_{\cS'}|\ZZ_{\ocS(\I')})$ by $H(Z_{\cS'}|Z_{\ocS(\I')})$
and $\oH(\ZZ_{\D'}|\YY_j,\ZZ_{\D(j)\setminus\D'})$
by $H(Z_{\D'}|Y_j,Z_{\D(j)\setminus\D'})$
to obtain a single-letter characterized {\em achievable} region.
By considering an extension of the problem (super problem \cite{ICC}) 
with zero rate auxiliary messages, which is analogous to introducing
auxiliary random variables,
we can obtain a potential extension of the single-letter
characterized achievable region,
where the specific cases are given in \cite{ITW13}\cite{ICC}\cite{CRNG-MULTI}.
In \cite{CRNG-MULTI},
we find a multi-letter characterized capacity region
by using the reduction technique introduced
in \cite[Problems 14.22-14.24]{CK11}\cite{H79}.
It should be noted here that
the characterization presented in this paper
provides a potentially larger achievable region
when channels and auxiliary sources are restricted to being stationary  and memoryless.
\end{rem}

\section{Joint Distribution Consistent with Access Structure}
Before proving Theorem \ref{thm:ROP=RIT},
we investigate the possible joint distribution
of random variables consistent with access structure~$\A$.

Let $\I'$ be a subset of $\I$.
We refer to $Z^n_{\bigcap_{i\in\I'}\cS(i)}$ 
as {\em common sources for Encoders $\I'$},
to $Z^n_{\cS(\I')}$ as {\em private sources for Encoders~$\I'$},
to $Z^n_{\bigcup_{\I''\in\fI:\I''\subset\I'}\cS(\I'')}$
as {\em private sources relevant to Encoders~$\I'$},
to $Z^n_{\lrB{\bigcap_{i\in\I'}\cS(i)}\cap\cS(\I')^{\complement}}$
as {\em public sources for Encoders $\I'$},
and to
$Z^n_{\lrB{\bigcap_{i\in\I'}\cS(i)}^{\complement}
  \cap\lrB{\bigcup_{\I''\in\fI:\I''\subset\I'}\cS(\I'')}^{\complement}}$
as {\em irrelevant sources to Encoders $\I'$}.

Here, let us consider the following problem.
Encoder $i$ has access to the set of random variables
$Z^n_{\cS(i)}\equiv\{Z^n_s\}_{s\in\cS(i)}$,
where $Z^n_s$ corresponds to message $M^{(n)}_s$
but random variables $\{Z^n_s\}_{s\in\cS}$ are allowed to 
be correlated on condition that
private sources for Encoders $\I'$
are allowed to be correlated with other sources
only through their public sources for all $\I'\in\fI$;
that is, they satisfy the following Markov relation
\begin{equation}
 Z^n_{\lrB{\bigcap_{i\in\I'}\cS(i)}^{\complement}
  \cap\lrB{\bigcup_{\I''\in\fI:\I''\subset\I'}\cS(\I'')}^{\complement}}
 \markov Z^n_{\lrB{\bigcap_{i\in\I'}\cS(i)}\cap\cS(\I')^{\complement}}
 \markov Z^n_{\cS(\I')}
 \label{eq:markov}
\end{equation}
for all $\I'\in\fI$.
It should be noted that
common sources for Encoders $\I'$ are allowed to be correlated.
Other encoders may have access to the public sources for Encoders $\I'$
but do not have access to the private sources relevant to Encoders $\I'$.
In this situation,
we specify the joint distribution of $\{Z^n_s\}_{s\in\cS}$.
The following lemma solves this problem.
\begin{lem}
\label{lem:joint}
The following two statements are equivalent.
\begin{itemize}
 \item 
 Joint source $Z^n_{\cS}$ satisfies
 (\ref{eq:markov}) for all $\I'\in\fI$.
 \item
 The joint distribution $\mu_{Z^n_{\cS}}$ of $Z^n_{\cS}$
 is given by (\ref{eq:jointZ}).
\end{itemize}
\end{lem}
\begin{IEEEproof}
Let $\ocS(\I')$ and $\dcS(\I')$ be defined by
(\ref{eq:oSI}) and (\ref{eq:dSI}), respectively.
Let $\{\I_k\}_{k=1}^{|\fI|}$
be defined at the end of Section \ref{sec:access}.

First, we show that
the first statement implies the second statement.
We have
\begin{align}
 \lrB{\bigcap_{i\in\I'}\cS(i)}\cap\cS(\I')^{\complement}
 &=
 \lrB{\cS(\I')\cup\ocS(\I')}\cap\cS(\I')^{\complement}
 \notag
 \\
 &=
 \lrB{\cS(\I')\cap\cS(\I')^{\complement}}
 \cup
 \lrB{\ocS(\I')\cap\cS(\I')^{\complement}}
 \notag
 \\
 &=
 \ocS(\I')\cap\cS(\I')^{\complement}
 \notag
 \\
 &=
 \lrB{\ocS(\I')\cap\cS(\I')^{\complement}}
 \cup\lrB{\ocS(\I')\cap\cS(\I')}
 \notag
 \\
 &=
 \ocS(\I')\cap\lrB{\cS(\I')^{\complement}\cup\cS(\I')}
 \notag
 \\
 &=
 \ocS(\I'),
 \label{eq:ZocSI'}
\end{align}
where the first equality comes from Lemma \ref{lem:cap-Si},
and the fourth equality comes from Lemma \ref{lem:disjoint-SIk-oSIk}.
From Lemma \ref{lem:cap-Si}, we have the fact that
\begin{align}
 \lrB{\bigcap_{i\in\I'}\cS(i)}^\complement
 \cap
 \lrB{\bigcup_{\I''\subseteq\I'}\cS(\I'')}^{\complement}
 &=
 \lrB{
  \lrB{\bigcap_{i\in\I'}\cS(i)}
  \cup
  \lrB{\bigcup_{\I''\subseteq\I'}\cS(\I'')}
 }^\complement
 \notag
 \\
 &=
 \lrB{
  \cS(\I')\cup\ocS(\I')
  \cup
  \lrB{\bigcup_{\I''\subseteq\I'}\cS(\I'')}
 }^\complement
 \notag
 \\
 &=
 \lrB{
  \ocS(\I')
  \cup
  \dcS(\I')
 }^{\complement}
 \notag
 \\
 &=
 \ocS(\I')^{\complement}
 \cap
 \dcS(\I')^{\complement}.
\end{align}
Then the condition (\ref{eq:markov}) is equivalent to
\begin{equation}
 Z^n_{
  \ocS(\I')^{\complement}
  \cap
  \dcS(\I')^{\complement}
 }
 \markov Z^n_{\ocS(\I')}
 \markov Z^n_{\cS(\I')}.
 \label{eq:markov-ocSI'}
\end{equation}
Then we have
\begin{align}
 \mu_{Z^n_{\cS}}
 (\zz_{\cS})
 &=
 \prod_{k=1}^{|\fI|}
 \mu_{Z^n_{\cS(\I_k)}|Z^n_{\bigcup_{k'=1}^{k-1}\cS(\I_{k'})}}
 (\zz_{\cS(\I_k)}|\zz_{\bigcup_{k'=1}^{k-1}\cS(\I_{k'})})
 \notag
 \\
 &=
 \prod_{k=1}^{|\fI|}
 \mu_{Z^n_{\cS(\I_k)}|Z^n_{\ocS(\I_k)}}
 (\zz_{\cS(\I_k)}|\zz_{\ocS(\I_k)})
 \notag
 \\
 &=
 \prod_{\I'\in\fI}
 \mu_{Z^n_{\cS(\I')}|Z^n_{\ocS(\I')}}
 (\zz_{\cS(\I')}|\zz_{\ocS(\I')}),
\end{align}
where
the first equality comes from Lemmas \ref{lem:union} and \ref{lem:disjoint},
the second equality comes from
(\ref{eq:ocSIk}), (\ref{eq:markov-ocSI'}),
and Lemma \ref{lem:k'<k}.

Next, we show that
the second statement implies the first statement.
Assume that
$\cS(\I')\cap\lrB{\bigcup_{\I'''\in\fI:\I'''\supsetneq\I''}\cS(\I''')}
\neq\emptyset$.
Since Lemmas \ref{lem:union} and \ref{lem:disjoint}
imply that
only one of $\I'=\I'''$
and $\cS(\I')\cap\cS(\I''')=\emptyset$
holds,
we have the fact that $\I'\supsetneq\I''$
and $\cS(\I'')\cap\ocS(\I')=\emptyset$.
Furthermore, $\I'\supsetneq\I''$ implies
$\cS(\I'')\subset\dcS(\I')$
and $\cS(\I'')\cap\dcS(\I')^{\complement}=\emptyset$.
Hence, we have the fact that
$\cS(\I'')\subset\ocS(\I')\cup\dcS(\I')^{\complement}$
implies $\cS(\I')\cap\ocS(\I'')
=\cS(\I')\cap\lrB{\bigcup_{\I'''\in\fI:\I'''\supsetneq\I''}\cS(\I''')}
=\emptyset$.
Since
\begin{align}
 \cS(\I')\cup\ocS(\I')\cup[\ocS(\I')\cup\dcS(\I')]^{\complement}
 &=
 \cS(\I')\cup\ocS(\I')\cup
 [\ocS(\I')^{\complement}\cap\dcS(\I')^{\complement}]
 \notag
 \\
 &=
 [\cS(\I')\cup\ocS(\I')\cup\ocS(\I')^{\complement}]
 \cap
 [\cS(\I')\cup\ocS(\I')\cup\dcS(\I')^{\complement}]
 \notag
 \\
 &=
 \cS(\I')\cup\ocS(\I')\cup\dcS(\I')^{\complement}
 \\
 \lrB{
  \cS(\I')\cup\ocS(\I')\cup[\ocS(\I')\cup\dcS(\I')]^{\complement}
 }^{\complement}
 &=
 \lrB{
  \cS(\I')\cup\ocS(\I')\cup\dcS(\I')^{\complement}
 }^{\complement}
 \notag
 \\
 &=
 \cS(\I')^{\complement}\cap\ocS(\I')^{\complement}\cap\dcS(\I'),
\end{align}
we have (\ref{eq:markov}) from
\begin{align}
 &
 \mu_{Z^n_{\cS(\I')\cup\ocS(\I')\cup[\ocS(\I')\cup\dcS(\I')]^{\complement}}}
 \lrsb{\zz_{\cS(\I')\cup\ocS(\I')\cup[\ocS(\I')\cup\dcS(\I')]^{\complement}}}
 \notag
 \\*
 &=
 \sum_{
  \zz_{
   \cS(\I')^{\complement}\cap\ocS(\I')^{\complement}\cap\dcS(\I')
  }
  \in\Z^n_{
   \cS(\I')^{\complement}\cap\ocS(\I')^{\complement}\cap\dcS(\I')
  }
 }
 \prod_{\I''\in\fI}
 \mu_{Z^n_{\cS(\I'')}|Z^n_{\ocS(\I'')}}
 \lrsb{\zz_{\cS(\I'')}|\zz_{\ocS(\I'')}}
 \notag
 \\
 &=
 \prod_{
   \I''\in\fI:
   \cS(\I'')\subset
   \cS(\I')\cup\ocS(\I')\cup\dcS(\I')^{\complement}
  }
 \mu_{Z^n_{\cS(\I'')}|Z^n_{\ocS(\I'')}}
 \lrsb{\zz_{\cS(\I'')}|\zz_{\ocS(\I'')}}
 \notag
 \\
 &=
 \mu_{Z^n_{\ocS(\I')\cup\dcS(\I')^{\complement}}}
 \lrsb{\zz_{\ocS(\I')\cup\dcS(\I')^{\complement}}}
 \mu_{Z^n_{\cS(\I')}|Z^n_{\ocS(\I')}}
 \lrsb{\zz_{\cS(\I')}|\zz_{\ocS(\I')}},
\end{align}
where the last equality comes from the fact that
$\cS(\I'')\subset\ocS(\I')\cup\dcS(\I')^{\complement}$
implies $\cS(\I')\cap\ocS(\I'')=\emptyset$.
\end{IEEEproof}

\section{Proof of the Converse}
\label{sec:converse}

In the following, we prove $\ROP\subset\RIT$.

Assume that $\{R_s\}_{s\in\cS}\in\ROP$ and let $r_s\equiv0$ for each $s\in\cS$.
Then there is a code
$\{(\{\Phi^{(n)}_i\}_{i\in\I},\{\Psi^{(n)}_j\}_{j\in\J})\}_{n=1}^{\infty}$
that satisfies (\ref{eq:ROP-rate}) and (\ref{eq:ROP-error})
for all $i\in\I$ and $j\in\J$.

For $j\in\J$ and $\D'\subset\D(j)$,
let $\Psi^{(n)}_{j,\D'}(Y^n_j)$
be the projection of $\Psi^{(n)}_j(Y^n_j)$ on $\M^{(n)}_{\D'}$.
Then we have
\begin{equation*}
 \limn P(\Psi^{(n)}_{j,\D'}(Y_j^n)\neq M^{(n)}_{\D'})=0
\end{equation*}
from (\ref{eq:ROP-error}).
From Lemmas~\ref{lem:oH(U|V)>oH(U|VW)} and \ref{lem:fano}
in Appendix \ref{sec:ispec},
we have
\begin{align}
 \oH(\MM_{\D'}|\YY_j,\MM_{\D(j)\setminus\D'})
 &\leq
 \oH(\MM_{\D'}|\YY_j)
 \notag
 \\
 &=0.
 \label{eq:HMgYM>=0}
\end{align}
From (\ref{eq:HMgYM>=0})
and $\oH(\MM_{\D'}|\YY_j,\MM_{\D(j)\setminus\D'})\geq 0$,
we have 
\begin{equation*}
 \oH(\MM_{\D'}|\YY_j,\MM_{\D(j)\setminus\D'})=0.
\end{equation*}
Then it is clear that
\begin{align}
 \sum_{s\in\D'}r_s\geq \oH(\MM_{\D'}|\YY_j,\MM_{\D(j)\setminus\D'})
 \label{eq:proof-converse-r}
\end{align}
for all $(j,\D')$ satisfying $j\in\J$ and $\emptyset\neq\D'\subset\D(j)$.

Assume that $\I'\in\fI$ and $\cS'\subset\cS(\I')$.
Since the distribution $\mu_{M^{(n)}_{\cS'}}$
is uniform on $\M^{(n)}_{\cS'}$,
we have the fact that
\begin{align}
 \frac1n\log_2\frac 1{\mu_{M^{(n)}_{\cS'}}(\mm_{\cS'})}
 &=
 \frac1n\log_2|\M^{(n)}_{\cS'}|
 \notag
 \\
 &\geq
 \liminfn\frac 1n \log_2|\M^{(n)}_{\cS'}|-\delta
\end{align}
for all $\mm_{\cS'}\in\M^{(n)}_{\cS'}$, $\delta>0$,
and all sufficiently large $n$.
This implies that
\begin{align}
 \limn
 \Prob\lrsb{
  \frac1n\log_2\frac 1{\mu_{M^{(n)}_{\cS'}}(M^{(n)}_{\cS'})}
  <\liminfn\frac 1n \log_2|\M^{(n)}_{\cS'}|-\delta
 }
 &=
 0.
 \label{eq:proof-converse-PM}
\end{align}
Let $\MM_{\cS'}\equiv\{M^{(n)}_{\cS'}\}_{n=1}^{\infty}$ be a general source.
Then we have 
\begin{align}
 \liminfn\frac 1n \log_2|\M^{(n)}_{\cS'}|-\delta
 &\leq
 \uH(\MM_{\cS'})
 \notag
 \\
 &=
 \uH(\MM_{\cS'}|\MM_{\ocS(\I')}),
 \label{eq:proof-converse-uHM}
\end{align}
where the inequality comes from (\ref{eq:proof-converse-PM})
and the definition of $\uH(\MM_{\cS'})$ in Appendix \ref{sec:ispec},
and the equality comes from
the fact that Lemma \ref{lem:disjoint-SIk-oSIk}
implies that $M^{(n)}_{\cS'}$ and $M^{(n)}_{\ocS(\I')}$ are independent.
We have
\begin{align}
 \sum_{s\in\cS'}
 [R_s+r_s]
 &=
 \sum_{s\in\cS'}
 R_s
 \notag
 \\
 &\leq
 \liminfn\frac{\log_2|\M^{(n)}_{\cS'}|}n
 \notag
 \\
 &\leq
 \uH(\MM_{\cS'}|\MM_{\ocS(\I')})+\delta,
\end{align}
where the equality comes from the fact that $r_s=0$ for all $s\in\cS$,
the first inequality comes from  (\ref{eq:ROP-rate}),
and the second inequality comes from  (\ref{eq:proof-converse-uHM}).
By letting $\delta\to0$, we have
\begin{align}
 \sum_{s\in\cS'}
 [R_s+r_s]
 &\leq
 \uH(\MM_{\cS'}|\MM_{\ocS(\I')})
 \label{eq:proof-converse-rR}
\end{align}
for all $(\I',\cS')$
satisfying $\I'\in\fI$ and $\emptyset\neq\cS'\subset\cS(\I')$.
Let $\ZZ_s\equiv\MM_s$ for each $s\in\cS$
and $\XX_i\equiv\{\Phi^{(n)}_i(M^{(n)}_{\cS(i)})\}_{n=1}^{\infty}$
for each $i\in\I$.
From Lemma \ref{lem:joint}
and the fact that $Z^n_{\cS}$ satisfies (\ref{eq:markov}) for all $\I'\in\fI$,
the joint distribution of $(Z^n_{\cS},X^n_{\I},Y_{\J}^n)$ is
allowed to be given by (\ref{eq:jointZ}) and (\ref{eq:jointXYZ}).
Then, from (\ref{eq:proof-converse-r}) and (\ref{eq:proof-converse-rR}),
we have $\{R_s\}_{s\in\cS}\in\RIT$, which implies $\ROP\subset\RIT$.
\hfill\IEEEQED

\section{Construction of Channel Code}
\label{sec:channel-code}

This section introduces a channel code based on the idea drawn from
\cite{CRNG}\cite{HASH}\cite{HASH-BC}\cite{HASH-MAC}\cite{CRNG-CHANNEL};
a similar idea is found in \cite[Theorem 14.3]{CK11}\cite{YAG12}.

For each $s\in\cS$, let us introduce a set $\C^{(n)}_s$
and two functions $f_s:\Z^n_s\to\C^{(n)}_s$ and $g_s:\Z^n_s\to\M^{(n)}_s$,
where the dependence of $f_s$ and $g_s$ on $n$ is omitted.
We can use sparse matrices as functions $f_s$ and  $g_s$ by assuming that
$\Z_s^n$, $\C^{(n)}_s$, and $\M^{(n)}_s$ are linear spaces
on the same finite field.
For a given $\cS'\subset\cS$ and $f_{\cS'}\equiv\{f_s\}_{s\in\cS'}$,
$g_{\cS'}\equiv\{g_s\}_{s\in\cS'}$, $\cc_{\cS'}\equiv\{\cc_s\}_{s\in\cS'}$,
$\mm_{\cS'}\equiv\{\mm_s\}_{s\in\cS'}$,
let
\begin{align*}
 \fC_{f_{\cS'}}(\cc_{\cS'})
 &\equiv
 \{\zz_{\cS'}: f_s(\zz_s)=\cc_s\ \text{for all}\ s\in\cS'\}
 \\
 \fC_{(f,g)_{\cS'}}(\cc_{\cS'},\mm_{\cS'})
 &\equiv
 \{\zz_{\cS'}: f_s(\zz_s)=\cc_s, g_s(\zz_s)=\mm_s\ \text{for all}\ s\in\cS'\},
\end{align*}
where $\zz_{\cS'}\equiv\{\zz_s\}_{s\in\cS'}$
and $(f,g)_{\cS'}(\zz)\equiv\{(f_s(\zz),g_s(\zz))\}_{s\in\cS'}$.
We define $\chi(\mathrm{S})$ as
\begin{gather*}
 \chi(\mathrm{S})
 \equiv
 \begin{cases}
  1
  &\text{if statement $\mathrm{S}$ is true}
  \\
  0
  &\text{if statement $\mathrm{S}$ is false}.
 \end{cases}
\end{gather*}

We fix two sets of functions $f_{\cS}$ and $g_{\cS}$,
and a set of vectors $\cc_{\cS}$
such that they are available for constructing encoders and decoders.
For each $i\in\I$,
Encoder $i$ uses $f_{\cS(i)}$, $g_{\cS(i)}$, and $\cc_{\cS(i)}$.
For each $j\in\J$,
Decoder $j$ uses $f_{\D(j)}$, $g_{\D(j)}$, and $\cc_{\D(j)}$.
We fix the probability distribution
$\mu_{Z^n_{\cS}}\equiv\{\mu_{Z^n_s}\}_{s\in\cS}$ given by (\ref{eq:jointZ})
and conditional probability distributions
$\{\mu_{X_i^n|Z^n_{\cS(i)}}\}_{i\in\I}$.

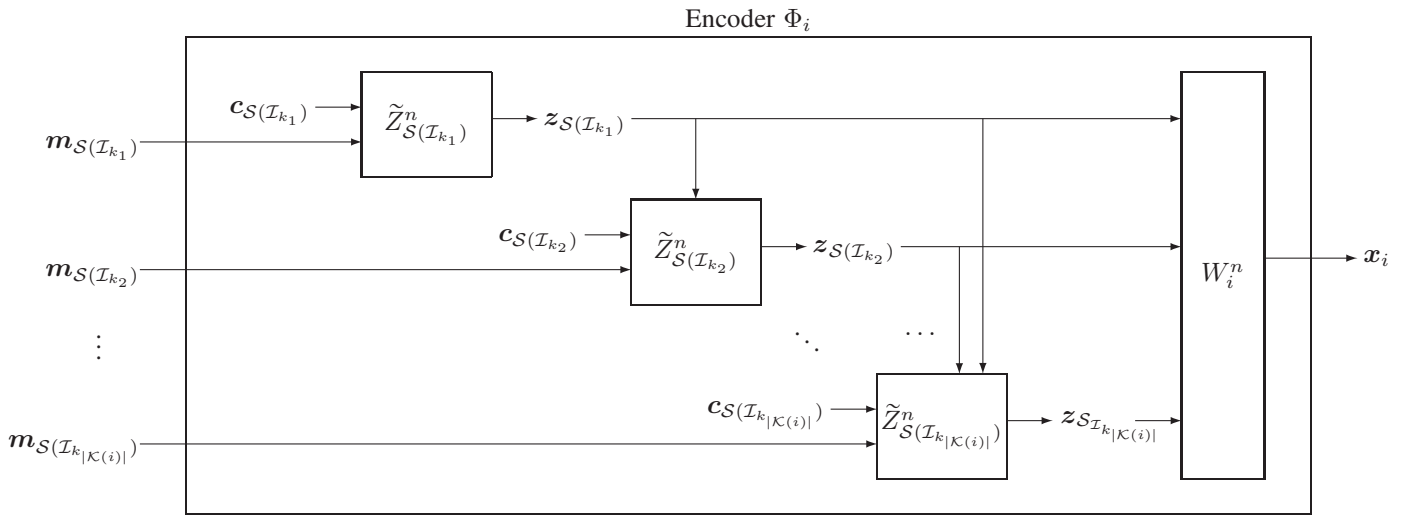
\begin{figure}
\begin{center}
 \unitlength 0.77mm
 \begin{picture}(240,95)(0,0)
  \put(126,85){\makebox(0,0){Encoder $\Phi_i$}}
  \put(22,63){\makebox(0,0)[r]{$\mm_{\cS(\I_{k_1})}$}}
  \put(22,41){\makebox(0,0)[r]{$\mm_{\cS(\I_{k_2})}$}}
  \put(15,30){\makebox(0,0){$\vdots$}}
  \put(22,11){\makebox(0,0)[r]{$\mm_{\cS(\I_{k_{|\K(i)|}})}$}}

  \put(51,69){\makebox(0,0)[r]{$\cc_{\cS(\I_{k_1})}$}}
  \put(52,70){\vector(1,0){8}}
  \put(97,47){\makebox(0,0)[r]{$\cc_{\cS(\I_{k_2})}$}}
  \put(98,48){\vector(1,0){8}}

  \put(139,17){\makebox(0,0)[r]{$\cc_{\cS(\I_{k_{|\K(i)|}})}$}}
  \put(140,18){\vector(1,0){8}}

  \put(22,64){\vector(1,0){38}}
  \put(22,42){\vector(1,0){84}}
  \put(22,12){\vector(1,0){126}}

  \put(60,58){\framebox(22,18){$\tZ^n_{\cS(\I_{k_1})}$}}
  \put(82,68){\vector(1,0){8}}
  \put(98,67){\makebox(0,0){$\zz_{\cS(\I_{k_1})}$}}
  \put(106,68){\vector(1,0){94}}
  \put(117,68){\vector(0,-1){14}}

  \put(106,36){\framebox(22,18){$\tZ^n_{\cS(\I_{k_2})}$}}
  \put(128,46){\vector(1,0){8}}
  \put(144,45){\makebox(0,0){$\zz_{\cS(\I_{k_2})}$}}
  \put(152,46){\vector(1,0){48}}

  \put(136,31){\makebox(0,0){$\ddots$}}

  \put(166,68){\vector(0,-1){44}}
  \put(162,46){\vector(0,-1){22}}
  \put(158,31){\makebox(0,0)[r]{$\ldots$}}

  \put(148,6){\framebox(22,18){$\tZ^n_{\cS(\I_{k_{|\K(i)|}})}$}}
  \put(170,16){\vector(1,0){8}}
  \put(188,15){\makebox(0,0){$\zz_{\cS_{\I_{k_{|\K(i)|}}}}$}}
  \put(192,16){\vector(1,0){8}}

  \put(200,6){\framebox(14,70){$W^n_i$}}
  \put(214,44){\vector(1,0){16}}
  \put(231,44){\makebox(0,0)[l]{$\xx_i$}}
  \put(30,0){\framebox(192,82){}}
 \end{picture}
\end{center}
\caption{Construction of Encoder $i$:
 It is assumed that
 $\K(i)\equiv\{k: i\in\I_k\}\equiv\{k_1,k_2,\ldots,k_{|\K(i)|}\}$
 satisfies $k_1<k_2<\cdots<k_{|\K(i)|}$.
 Arrows from $\zz_{\cS(I_{k'})}$ to $\tZ_{\cS(I_k)}$
 are ignored when $\I_k\subsetneq\I_{k'}$ is not satisfied.}
\label{fig:encoder}
\end{figure}

We define a constrained-random-number generator for encoder use.
For given $\I'\in\fI$,
$\zz_{\ocS(\I')}$, $\cc_{\cS(\I')}$, and $\mm_{\cS(\I')}$,
let $\tZ^n_{\cS(\I')}$ be a random variable corresponding to the distribution
\begin{align}
 &
 \mu_{\tZ^n_{\cS(\I')}|\tZ^n_{\ocS(\I')}C^{(n)}_{\cS(\I')}M^{(n)}_{\cS(\I')}}
 (\zz_{\cS(\I')}|\zz_{\ocS(\I')},\cc_{\cS(\I')},\mm_{\cS(\I')})
 \notag
 \\*
 &\equiv
 \frac{
  \mu_{Z^n_{\cS(\I')}|Z^n_{\ocS(\I')}}(\zz_{\cS(\I')}|\zz_{\ocS(\I')})
  \chi(
   \zz_{\cS(\I')}\in\fC_{(f,g)_{\cS(\I')}}(\cc_{\cS(\I')},\mm_{\cS(\I')})
  )
 }{
  \mu_{Z^n_{\cS(\I')}|Z^n_{\ocS(\I')}}
  (
   \fC_{(f,g)_{\cS(\I')}}(\cc_{\cS(\I')},\mm_{\cS(\I')})
   |\zz_{\ocS(\I')}
  )
 },
 \label{eq:tZI}
\end{align}
where $\{\ocS(\I')\}_{I'\in\fI}$
is obtained before encoding by employing Algorithm \ref{alg:ocSIk}.
We assume that the constrained-random number generator outputs the same $\zz_{\cS(\I')}$ to all encoders that have access to message $\mm_{\cS(\I')}$ for a given $\I'\in\fI$.
Then Encoder $i$ generates $\zz_{\cS(i)}$ by using Algorithm \ref{alg:zSi}
based on Lemma \ref{lem:unionSIk=Si} and (\ref{eq:jointZ}),
where (\ref{eq:ocSIk}) implies that
Encoder $i$ has obtained $\zz_{\ocS(\I_k)}$
at Line 2 of Algorithm \ref{alg:zSi}.
We define encoding function  $\Phi^{(n)}_i:\M^{(n)}_{\cS(i)}\to\X_i^n$ as
\begin{equation*}
 \Phi^{(n)}_i(\mm_{\cS(i)})
 \equiv
 W^n_i(\tZ^n_{\cS(i)}),
\end{equation*}
where the encoder claims an error when
$\mu_{Z_{\cS(\I')}^n|Z_{\ocS(\I')}^n}
  (\fC_{(f,g)_{\cS(\I')}}(\cc_{\cS(\I')},\mm_{\cS(\I')})|\zz_{\ocS(\I')})=0$
and $W^n_i$ is the channel corresponding to
the conditional probability distribution $\mu_{X^n_i|Z^n_{\cS(i)}}$.
The flow of vectors is illustrated in Fig.\ \ref{fig:encoder}.

\begin{rem}
By using the interval algorithm introduced in \cite{CRNG},
encoders can share the same output of
a given constrained-random-number generator
by sharing a fixed real number belonging to $[0,1]$.
\end{rem}

\begin{algorithm}[t]
 \caption{Construction of $\{\ocS(\I')\}_{I'\in\fI}$}
 \hspace*{\algorithmicindent}\textbf{Input:}
 List $\fI\equiv\{\I_k\}_{k=1}^{|\fI|}$,
 which is sorted so that $\I_k\subsetneq\I_{k'}$ implies $k'<k$
 for all $k,k'\in\{1,2,\ldots,|\fI|\}$.
 \\
 \hspace*{\algorithmicindent}\phantom{\textbf{Input:}}
 List $\{\cS(\I_k)\}_{k=1}^{|\fI|}$.
 \\
 \hspace*{\algorithmicindent}\textbf{Output:}
 List $\{\ocS(\I_k)\}_{k=1}^{|\fI|}$.
 \label{alg:ocSIk}
 \begin{algorithmic}[1]
  \For{$k\in\{1,\ldots,|\fI|\}$}
    \State $\ocS(\I_k)\leftarrow\emptyset$
    \For{$k'\in\{1,\ldots,k-1\}$}
      \IIf{$\I_{k'}\supsetneq\I_k$}
        $\ocS(\I_k)\leftarrow\ocS(\I_k)\cup\cS(\I_{k'})$.
    \EndFor
  \EndFor
 \end{algorithmic}
\end{algorithm}

\begin{algorithm}[t]
 \caption{Generation of $\zz_{\cS(i)}$}
 \hspace*{\algorithmicindent}\textbf{Input:}
 Lists $\{\cS(\I_k)\}_{k=1}^{|\fI|}$, $\{\ocS(\I_k)\}_{k=1}^{|\fI|}$,
 $\cc_{\cS(i)}\equiv\{\cc_s\}_{s\in\cS(i)}$,
 and $\mm_{\cS(i)}\equiv\{\mm_s\}_{s\in\cS(i)}$.
 \\
 \hspace*{\algorithmicindent}\textbf{Output:}
 Vectors $\zz_{\cS(i)}\equiv\{\zz_s\}_{s\in\cS(i)}$.
 \label{alg:zSi}
 \begin{algorithmic}[1]
  \For{$k\in\{1,\ldots,|\fI|\}$}
   \IIf{$i\in\I_k$}
   generate
   $\zz_{\cS(\I_k)}$
   subject to the distribution
   $\mu_{\tZ^n_{\cS(\I_k)}|Z^n_{\ocS(\I_k)}C^{(n)}_{\cS(\I_k)}M^{(n)}_{\cS(\I_k)}}
   (\cdot|\zz_{\ocS(\I_k)},\cc_{\cS(\I_k)},\mm_{\cS(\I_k)})$.
  \EndFor
 \end{algorithmic}
\end{algorithm}

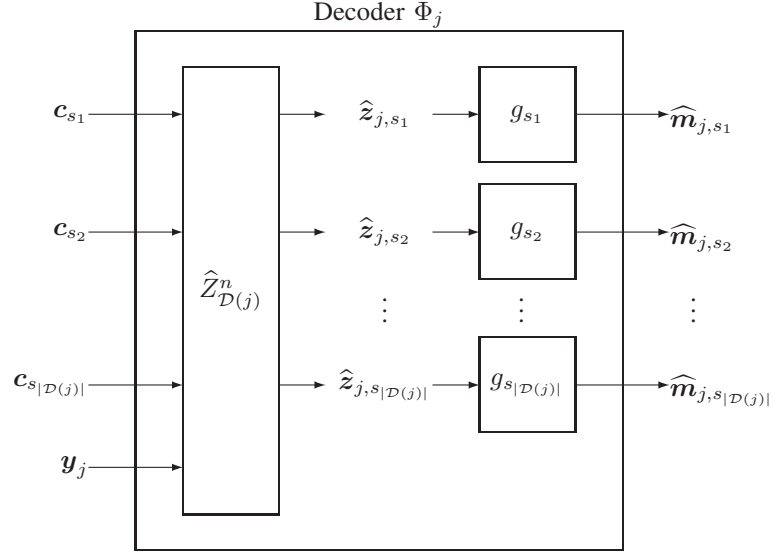
\begin{figure}
\begin{center}
 \unitlength 0.78mm
 \begin{picture}(126,100)(0,4)
  \put(67,95){\makebox(0,0){Decoder $\Phi_j$}}
  \put(116,77){\makebox(0,0)[l]{$\hmm_{j,s_1}$}}
  \put(116,57){\makebox(0,0)[l]{$\hmm_{j,s_2}$}}
  \put(120,46){\makebox(0,0){$\vdots$}}
  \put(116,31){\makebox(0,0)[l]{$\hmm_{j,s_{|\D(j)|}}$}}
  \put(100,78){\vector(1,0){16}}
  \put(100,58){\vector(1,0){16}}
  \put(100,32){\vector(1,0){16}}
  \put(84,70){\framebox(16,16){$g_{s_1}$}}
  \put(84,50){\framebox(16,16){$g_{s_2}$}}
  \put(91,46){\makebox(0,0){$\vdots$}}
  \put(84,24){\framebox(16,16){$g_{s_{|\D(j)|}}$}}
  \put(50,78){\vector(1,0){8}}
  \put(50,58){\vector(1,0){8}}
  \put(50,32){\vector(1,0){8}}
  \put(68,78){\makebox(0,0){$\hzz_{j,s_1}$}}
  \put(68,58){\makebox(0,0){$\hzz_{j,s_2}$}}
  \put(68,46){\makebox(0,0){$\vdots$}}
  \put(68,32){\makebox(0,0){$\hzz_{j,s_{|\D(j)|}}$}}
  \put(76,78){\vector(1,0){8}}
  \put(76,58){\vector(1,0){8}}
  \put(76,32){\vector(1,0){8}}
  \put(34,10){\framebox(16,76){$\hZ^n_{\D(j)}$}}
  \put(18,78){\vector(1,0){16}}
  \put(18,58){\vector(1,0){16}}
  \put(18,32){\vector(1,0){16}}
  \put(18,18){\vector(1,0){16}}
  \put(18,78){\makebox(0,0)[r]{$\cc_{s_1}$}}
  \put(18,58){\makebox(0,0)[r]{$\cc_{s_2}$}}
  \put(18,32){\makebox(0,0)[r]{$\cc_{s_{|\D(j)|}}$}}
  \put(18,18){\makebox(0,0)[r]{$\yy_j$}}
  \put(26,4){\framebox(82,88){}}
 \end{picture}
\end{center}
\caption{Construction of Decoder $j$:
 It is assumed that $\D(j)\equiv\{s_1,s_2,\ldots,s_{|\D(j)|}\}$.}
\label{fig:decoder}
\end{figure}

We define a constrained-random-number generator used by Decoder $j$.
For each $j\in\J$, Decoder $j$ generates
$\hzz_{\D(j)}\equiv\{\hzz_s\}_{s\in\D(j)}$
by using a constrained-random-number generator whose distribution is given as
\begin{align}
 \mu_{\hZ_{\D(j)}^n|C^{(n)}_{\D(j)}Y_j^n}(\hzz_{\D(j)}|\cc_{\D(j)},\yy_j)
 &\equiv
 \frac{
  \mu_{Z_{\D(j)}^n|Y_j^n}(\hzz_{\D(j)}|\yy_j)
  \chi(f_{\D(j)}(\hzz_{\D(j)})=\cc_{\D_j})
 }{
  \mu_{Z_{\D(j)}^n|Y_j^n}(\fC_{f_{\D(j)}}(\cc_{\D(j)})|\yy_j)
 }
 \label{eq:decoder}
\end{align}
for given vector $\cc_{\D(j)}$ and side information $\yy_j\in\Y_j^n$,
where $f_{\D(j)}(\hzz_{\D(j)})\equiv\lrb{f_s(\hzz_s)}_{s\in\D(j)}$.
We define the decoding function $\Psi^{(n)}_j:\Y_j^n\to\M^{(n)}_{\D(j)}$ as
\begin{equation*}
 \Psi^{(n)}_j(\yy_j)
 \equiv
 \{g_s(\hzz_{j,s})\}_{s\in\D(j)}.
\end{equation*}
The flow of vectors is illustrated in Fig.\ \ref{fig:decoder}.
It should be noted here that $\hZ^n_{\D(j)}$ is analogous
to the decoder reproducing the output
$\zz_{\D(j)}\equiv\{\zz_s\}_{s\in\D(j)}$ of correlated sources,
where $\cc_s\equiv f_s(\zz_s)$ corresponds
to the codeword by using encoding function $f_s$.

It should be noted that, for sources that are memoryless,
the tractable approximation algorithms
for a constrained-random-number generator
summarized in~\cite{SDECODING} can be used;
the maximum a posteriori probability decoder
is optimal but may be intractable.

Let
\begin{align*}
 r_s
 &\equiv
 \frac{\log|\C^{(n)}_s|}n
 \\
 R_s
 &\equiv
 \frac{\log|\M^{(n)}_s|}n,
\end{align*}
where $R_s$ represents the rate of Message $s$.
Let $\hM^{(n)}_{\D(j)}\equiv \Psi^{(n)}_j(Y^n_j)$
and $\Error(f_{\cS},g_{\cS},\cc_{\cS})$ be the error probability defined as
\begin{equation}
 \Error(f_{\cS},g_{\cS},\cc_{\cS})
 \equiv
 \Prob\lrsb{
  \begin{aligned}
   &\hM^{(n)}_{j,s}\neq M^{(n)}_s\ \text{for some}\ j\in\J, s\in\D(j),
   \\
   &\text{or}
   \ \mu_{Z_{\cS(\I')}^n|Z_{\ocS(\I')}^n}
   (\fC_{(f,g)_{\cS(\I')}}(\cc_{\cS(\I')},M^{(n)}_{\cS(\I')})|Z^n_{\ocS(\I')})=0
   \ \text{for some}\ \I'\in\fI
  \end{aligned}
 }.
 \label{eq:error}
\end{equation}
The following theorem,
implies the achievability part, $\RIT\subset\ROP$,
of Theorem \ref{thm:ROP=RIT}.
The proof is given in
Section~\ref{sec:proof-channel}.
\begin{thm}
\label{thm:channel}
For a given access structure $(\cS,\I,\A)$,
let us define $\fI$, $\cS(\I')$, and $\ocS(\I')$
as described in Section~\ref{sec:access}.
Let us assume that
$\{(r_s,R_s)\}_{s\in\cS}$ satisfies
\begin{align}
 R_s
 &\geq
 0
 \label{eq:rate-positive}
 \\
 \sum_{s\in\cS'}[R_s+r_s]
 &<
 \uH(\ZZ_{\cS'}|\ZZ_{\ocS(\I')})
 \label{eq:rate-encoder}
 \\
 \sum_{s\in\D'}r_s
 &>
 \oH(\ZZ_{\D'}|\YY_j,\ZZ_{\D(j)\setminus\D'})
 \label{eq:rate-decoder}
\end{align}
for all $(\I',\cS',j,\D')$ satisfying
$\I'\in\fI$, $\emptyset\neq\cS'\subset\cS(\I')$,
$j\in\J$, and
$\emptyset\neq\D'\subset\D(j)$,
where the joint distribution of $(Z^n_{\cS},X^n_{\I},Y^n_{\J})$
is given by (\ref{eq:jointZ}) and (\ref{eq:jointXYZ}).
Then for all $\delta>0$ and all sufficiently large $n$
there are functions (sparse matrices) $f_{\cS}$, $g_{\cS}$,
and a set of vectors $\cc_{\cS}$ such that
$\Error(f_{\cS},g_{\cS},\cc_{\cS})\leq\delta$.
\end{thm}

\begin{rem}
It should be noted that
for specific $\mu_{Z^n_{\cS}X_{\I}^nY_{\J}^n}$ and $\{R_s\}_{s\in\cS}$
we can find $\{r_s\}_{s\in\cS}$
satisfying (\ref{eq:rate-encoder}) and (\ref{eq:rate-decoder})
by employing linear programming whenever they exist.
\end{rem}

\section{Proof of Theorem~\ref{thm:channel}}
\label{sec:proof-channel}

In the following,
we omit the dependence of $C$, $M$, $X$, $Y$, and $Z$ on $n$.

Let us assume that
ensembles $(\F_s,p_{F_s})$ and $(\G_s,p_{G_s})$,
where their dependence on $n$ is omitted,
have the hash property ((\ref{eq:hash}) described in Appendix \ref{sec:hash})
for every $s\in\cS$.
For each $s\in\cS$, let
\begin{align*}
 \C_s
 &\equiv
 \im\F_s
 \\
 &\equiv
 \bigcup_{f\in\F}\{f(\zz): \zz\in\Z^n\}
 \\
 \M_s
 &\equiv
 \im\G_s
 \\
 &\equiv
 \bigcup_{g\in\G}\{g(\zz): \zz\in\Z^n\},
\end{align*}
where we omit the dependence of $\C_s$ and $\M_s$ on $n$.
We use the fact without notice that
\begin{equation*}
 \{\fC_{(f,g)_{\cS'}}(\cc_{\cS'},\mm_{\cS'})\}_{\cc_{\cS'}
  \in\C_{\cS'},\mm_{\cS'}\in\M_{\cS'}}
\end{equation*}
forms a partition of $\Z_{\cS'}^n$ for a given $\cS'\subset\cS$.
For a given $k\in\{1,2,\ldots,|\fI|\}$,
let us define
\begin{align*}
 \cS_k
 &\equiv 
 \cS(\I_k)
 \\
 \ocS_k
 &\equiv
 \ocS(\I_k)
 \\
 \cS^k
 &\equiv
 \bigcup_{k'=1}^k
 \cS(\I_{k'}),
\end{align*}
where we $\{\I_k\}_{k=1}^{|\fI|}$ is defined
at the end of Section \ref{sec:access}.
We use the fact without notice that
$\{\cS_k\}_{k=1}^{|\fI|}$ forms a partition of $\cS$,
where it is shown by Lemmas \ref{lem:union} and \ref{lem:disjoint}.
We use the fact without notice that $\cS_k$ and $\ocS_k$ are disjoint,
where it is shown by Lemma~\ref{lem:disjoint-SIk-oSIk}.
Since (\ref{eq:ocSIk}) implies
\begin{align}
 \ocS_k
 &\subset
 \bigcup_{k'=1}^{k-1}
 \cS(\I_{k'})
 \notag
 \\
 &=
 \cS^{k-1},
\end{align}
we have
\begin{equation}
 \mu_{Z_{\cS_k}|Z_{\cS^{k-1}}}
 =
 \mu_{Z_{\cS_k}|Z_{\ocS_k}}
 \label{eq:mu_ZSk|ZoSk}
\end{equation}
from (\ref{eq:jointZ}) and Lemmas \ref{lem:k'<k} and \ref{lem:joint}.

Let 
\begin{align*}
 \E((f,g)_{\cS_k},\cc_{\cS_k})
 &\equiv
 \lrb{
  \mm_{\cS_k}:
  \mu_{Z_{\cS_k}|Z_{\ocS_k}}
  (
   \fC_{(f,g)_{\cS_k}}(\cc_{\cS_k},\mm_{\cS_k})
   |\zz_{\ocS_k}
  )=0
  \ \text{for some}
  \ \zz_{\ocS_k}\in\Z^n_{\ocS_k}
 }
 \\
 \E((f,g)_{\cS},\cc_{\cS})
 &\equiv
 \lrb{
  \mm_{\cS}:
  \mm_{\cS_k}\in\E((f,g)_{\cS_k},\cc_{\cS_k})
  \ \text{for some}\ k\in\{1,\ldots,|\fI|\}
 }
 \\
 \E(g_{\cS},\mm_{\cS})
 &\equiv
 \lrb{
  \hzz_{\D(\J)}:
  g_s(\hzz_{j,s})\neq \mm_s
  \ \text{for some}\ j\in\J, s\in\D(j)
 },
\end{align*}
where $\hzz_{\D(\J)}\equiv\{\hzz_{j,s}\}_{j\in\J,s\in\D(j)}$.
It follows that error probability 
$\Error(f_{\cS},g_{\cS},\cc_{\cS})$ is evaluated as
\begin{align}
 \Error(f_{\cS},g_{\cS},\cc_{\cS})
 &\leq
 \sum_{
  \mm_{\cS}\in\E((f,g)_{\cS},\cc_{\cS})
 }
 \prod_{s\in\cS}
 \frac1{
  |\M_s|
 }
 \notag
 \\*
 &\quad
 +
 \sum_{\substack{
   \mm_{\cS}\notin\E((f,g)_{\cS},\cc_{\cS}),
   \zz_{\cS}\in\fC_{(f,g)_{\cS}}(\cc_{\cS},\mm_{\cS}),
   \xx_{\I}\in\X^n_{\I},
   \\
   \yy_{\J}\in\Y^n_{\J},
   \hzz_{\D(\J)}\in\E(g_{\cS},\mm_{\cS})
 }}
 \lrB{
  \prod_{j\in\J}
  \mu_{\hZ_{\D(j)}|C_{\D(j)Y_j}}(\hzz_{\D(j)}|\cc_{\D(j)},\yy_j)
 }
 \mu_{Y_{\J}|X_{\I}}(\yy_{\J}|\xx_{\I})
 \notag
 \\*
 &\qquad\cdot
 \lrB{
  \prod_{i\in\I}
  \mu_{X_i|Z_{\cS(i)}}(\xx_i|\zz_{\cS(i)})
 }
 \lrB{
  \prod_{k=1}^{|\fI|}
  \mu_{\tZ_{\cS_k}|\tZ_{\ocS_k}C_{\cS_k}M_{\cS_k}}(
   \zz_{\cS_k}
   |\zz_{\ocS_k},\cc_{\cS_k},\mm_{\cS_k}
  )
 }
 \lrB{
  \prod_{s\in\cS}\frac1{|\M_s|}
 },
 \label{eq:channel-error}
\end{align}
where the first term on the right hand side
corresponds to the encoding error probability
and the second term on the right hand side
corresponds to the decoding error probability.

By using the union bound, the first term on the right hand side of the equality
in (\ref{eq:channel-error}) is evaluated as
\begin{equation*}
 \text{[the first term of (\ref{eq:channel-error})]}
 \leq
 \sum_{k=1}^{|\fI|}
 \sum_{
   \mm_{\cS_k}\in\E((f,g)_{\cS_k},\cc_{\cS_k})
 }
 \frac1{|\M_{\cS_k}|}.
 \label{eq:channel-error-1}
\end{equation*}

The second term on the right hand side of the equality
in (\ref{eq:channel-error}) is evaluated as
\begin{align}
 &
 \text{[the second term of (\ref{eq:channel-error})]}
 \notag
 \\*
 &=
 \sum_{\substack{
   \mm_{\cS}\notin\E((f,g)_{\cS},\cc_{\cS}),
   \zz_{\cS}\in\fC_{(f,g)_{\cS}}(\cc_{\cS},\mm_{\cS}),
   \\
   \yy_{\J}\in\Y^n_{\J},
   \hzz_{\D(\J)}\in\E(g_{\cS},\mm_{\cS})
 }}
 \frac{
  \mu_{\hZ_{\D(\J)}|C_{\cS}Y_{\J}}(\hzz_{\D(\J)}|\cc_{\cS},\yy_{\J})
  \mu_{Z_{\cS}\YY_{\J}}(\zz_{\cS},\yy_{\J})
 }{
  \prod_{k=1}^{|\fI|}
  \mu_{Z_{\cS_k}|Z_{\cS^{k-1}}}(
    \fC_{(f,g)_{\cS_k}}(\cc_{\cS_k},\mm_{\cS_k})
    |\zz_{\cS^{k-1}}
   )
   |\M_{\cS_k}|
 }
 \notag
 \\
 &\leq
 \sum_{\substack{
   \mm_{\cS}\notin\E((f,g)_{\cS},\cc_{\cS}),
   \zz_{\cS}\in\fC_{(f,g)_{\cS}}(\cc_{\cS},\mm_{\cS}),
   \yy_{\J}\in\Y^n_{\J},
   \hzz_{\D(\J)}\in\E(g_{\cS},\mm_{\cS})
 }}
 \mu_{\hZ_{\D(\J)}|C_{\cS}Y_{\J}}(\hzz_{\D(\J)}|\cc_{\cS},\yy_{\J})
 \mu_{Z_{\cS}\YY_{\J}}(\zz_{\cS},\yy_{\J})
 \notag
 \\*
 &\qquad\cdot
 |\C_{\cS}|
 \lrB{
  \lrbar{
   \frac1{
    \prod_{k=1}^{|\fI|}
    \mu_{Z_{\cS_k}|Z_{\cS^{k-1}}}(
     \fC_{(f,g)_{\cS_k}}(\cc_{\cS_k},\mm_{\cS_k})
     |\zz_{\cS^{k-1}})
    |\C_{\cS_k}|
    |\M_{\cS_k}|
   }
   -1
  }
  +
  1
 }
 \notag
 \\
 &\leq
 |\C_{\cS}|
 \sum_{\substack{
   \mm_{\cS}\notin\E((f,g)_{\cS},\cc_{\cS}),
   \zz_{\cS}\in\fC_{(f,g)_{\cS}}(\cc_{\cS},\mm_{\cS})
 }}
 \mu_{Z_{\cS}}(\zz_{\cS})
 \lrbar{
  \prod_{k=1}^{|\fI|}
  \frac1{
   \mu_{Z_{\cS_k}|Z_{\cS^{k-1}}}(
    \fC_{(f,g)_{\cS_k}}(\cc_{\cS_k},\mm_{\cS_k})
    |\zz_{\cS^{k-1}})
   |\C_{\cS_k}|
   |\M_{\cS_k}|
  }
  -1
 }
 \notag
 \\*
 &\quad
 +
 \sum_{j\in\J}
 |\C_{\cS}|
 \sum_{\substack{
   \mm_{\cS}\in\M_{\cS},
   \zz_{\cS}\in\fC_{(f,g)_{\cS}}(\cc_{\cS},\mm_{\cS}),
   \yy_{\J}\in\Y^n_{\J},
   \hzz_{\D(\J)}\in\Z^n_{\D(\J)}:
   \\
   \hzz_{\D(j)}\neq \zz_{\D(j)}
 }}
 \mu_{\hZ_{\D(\J)}|C_{\cS}Y_{\J}}(\hzz_{\D(\J)}|\cc_{\cS},\yy_{\J})
 \mu_{Z_{\cS}\YY_{\J}}(\zz_{\cS},\yy_{\J}),
 \label{eq:channel-error-2}
\end{align}
where the equality comes from (\ref{eq:tZI}), (\ref{eq:mu_ZSk|ZoSk}),
and the relations
\begin{align*}
 \mu_{\hZ_{\D(\J)}|C_{\cS}Y_{\J}}(\hzz_{\D(\J)}|\cc_{\cS},\yy_{\J})
 &\equiv
 \prod_{j\in\J}
 \mu_{\hZ_{\D(j)}|C_{\D(j)}Y_j}(\hzz_{\D(j)}|\cc_{\D(j)},\yy_j)
 \\
 \mu_{Z_{\cS}\YY_{\J}}(\zz_{\cS},\yy_{\J})
 &=
 \sum_{\xx_{\I}\in\X^n_{\I}}
 \mu_{Y_{\J}|X_{\I}}(\yy_{\J}|\xx_{\I})
 \lrB{
  \prod_{i\in\I}
  \mu_{X_i|Z_{\cS(i)}}(\xx_i|\zz_{\cS(i)})
 }
 \lrB{
  \prod_{k=1}^{|\fI|}
  \mu_{Z_{\cS_k}|Z_{\ocS_k}}(\zz_{\cS_k}|\zz_{\ocS_k})
 }.
\end{align*}
The first inequality comes from the triangular inequality,
and the second inequality comes from the relation
\begin{equation*}
 \sum_{
  \yy_{\J}\in\Y^n_{\J},
  \hzz_{\D(\J)}\in\E(g_{\cS},\mm_{\cS})
 }
 \mu_{\hZ_{\D(\J)}|C_{\cS}Y_{\J}}(\hzz_{\D(\J)}|\cc_{\cS},\yy_{\J})
 \mu_{Z_{\cS}\YY_{\J}}(\zz_{\cS},\yy_{\J})
 \leq
 \mu_{Z_{\cS}}(\zz_{\cS})
\end{equation*}
and the union bound with the fact that
$\hzz_{\D(\J)}\in\E(g_{\cS},\mm_{\cS})$
implies $\hzz_{\D(j)}\neq\zz_{\D(j)}$ for some $j\in\J$.
The first term on the right hand side of (\ref{eq:channel-error-2})
is evaluated as
\begin{align}
 &
 [\text{the first term of (\ref{eq:channel-error-2})}]
 \notag
 \\*
 &\leq
 |\C_{\cS}|
 \sum_{\substack{
   \mm_{\cS}\in\M_{\cS},
   \zz_{\cS}\in\fC_{(f,g)_{\cS}}(\cc_{\cS},\mm_{\cS}):
   \\
   \mm_{\cS_k}\notin\E((f,g)_{\cS_k},\cc_{\cS_k})
   \ \text{for all}\ k\in\{1,\ldots,|\fI|\}
 }}
 \mu_{Z_{\cS}}(\zz_{\cS})
 \sum_{k=1}^{|\fI|}
 \lrbar{
  \frac1{
   \mu_{Z_{\cS_k}|Z_{\cS^{k-1}}}(
    \fC_{(f,g)_{\cS_k}}(\cc_{\cS_k},\mm_{\cS_k})
    |\zz_{\cS^{k-1}})
   |\C_{\cS_k}|
   |\M_{\cS_k}|
  }
  -1
 }
 \notag
 \\*
 &\quad\cdot
 \prod_{k'=k+1}^{|\fI|}
 \frac1{
  \mu_{Z_{\cS_{k'}}|Z_{\cS^{k'-1}}}(
   \fC_{(f,g)_{\cS_{k'}}}(\cc_{\cS_{k'}},\mm_{\cS_{k'}})
   |\zz_{\cS^{k'-1}})
  |\C_{\cS_{k'}}|
  |\M_{\cS_{k'}}|
 }
 \notag
 \\
 &=
 \sum_{k=1}^{|\fI|}
 |\C_{\cS}|
 \sum_{\substack{
   \mm_{\cS^{k-1}}\in\M_{\cS^{k-1}},
   \zz_{\cS^{k-1}}\in\fC_{(f,g)_{\cS^{k-1}}}(\cc_{\cS^{k-1}},\mm_{\cS^{k-1}}):
   \\
   \mm_{\cS_{k'}}\notin\E((f,g)_{\cS_{k'}},\cc_{\cS_{k'}})
   \ \text{for all}\ k'\in\{1,\ldots,k-1\}
 }}
 \mu_{Z_{\cS^{k-1}}}(\zz_{\cS^{k-1}})
 \notag
 \\*
 &\quad\cdot
 \sum_{\substack{
   \mm_{\cS_k}\notin\E((f,g)_{\cS_k},\cc_{\cS_k})
   \\
   \zz_{\cS_k}\in\fC_{(f,g)_{\cS_k}}(\cc_{\cS_k},\mm_{\cS_k}),
 }}
 \mu_{Z_{\cS_k}|Z_{\cS^{k-1}}}
 (\zz_{\cS_k}|\zz_{\cS^{k-1}})
 \lrbar{
  \frac1{
   \mu_{Z_{\cS_k}|Z_{\cS^{k-1}}}(
    \fC_{(f,g)_{\cS_k}}(\cc_{\cS_k},\mm_{\cS_k})
    |\zz_{\cS^{k-1}})
   |\C_{\cS_k}|
   |\M_{\cS_k}|
  }
  -1
 }
 \notag
 \\*
 &\quad\cdot
 \sum_{\substack{
   \mm_{\cS_{k+1}}\notin\E((f,g)_{\cS_{k+1}},\cc_{\cS_{k+1}})
   \\
   \zz_{\cS_{k+1}}
   \in\fC_{(f,g)_{\cS_{k+1}}}(\cc_{\cS_{k+1}},\mm_{\cS_{k+1}})
 }}
 \frac{
  \mu_{Z_{\cS_{k+1}}|Z_{\cS^k}}(
   \zz_{\cS_{k+1}}|\zz_{\cS^k}
  )
 }{
  \mu_{Z_{\cS_{k+1}}|Z_{\cS^k}}(
   \fC_{(f,g)_{\cS_{k+1}}}(\cc_{\cS_{k+1}},\mm_{\cS_{k+1}})
   |\zz_{\cS^k})
  |\C_{\cS_{k+1}}|
  |\M_{\cS_{k+1}}|
 }
 \notag
 \\*
 &\ \quad\vdots
 \notag
 \\*
 &\quad\cdot
 \sum_{\substack{
   \mm_{\cS(|\fI|)}\notin\E((f,g)_{\cS_{|\fI|}},\cc_{\cS_{|\fI|}})
   \\
   \zz_{\cS_{|\fI|}}
   \in\fC_{(f,g)_{\cS_{|\fI|}}}(\cc_{\cS_{|\fI|}},\mm_{\cS_{|\fI|}})
 }}
 \frac{
  \mu_{Z_{\cS_{|\fI|}}|Z_{\cS^{|\fI|-1}}}(
   \zz_{\cS_{|\fI|}}|\zz_{\cS^{|\fI|-1}}
  )
 }{
  \mu_{Z_{\cS_{|\fI|}}|Z_{\cS^{|\fI|-1}}}(
   \fC_{(f,g)_{\cS_{|\fI|}}}(\cc_{\cS_{|\fI|}},\mm_{\cS_{|\fI|}})
   |\zz_{\cS^{|\fI|-1}})
  |\C_{\cS_{|\fI|}}|
  |\M_{\cS_{|\fI|}}|
 }
 \notag
 \\
 &\leq
 \sum_{k=1}^{|\fI|}
 |\C_{\cS}|
 \sum_{\substack{
   \zz_{\cS^{k-1}}
   \in\fC_{f_{\cS^{k-1}}}(\cc_{\cS^{k-1}})
 }}
 \mu_{Z_{\cS^{k-1}}}(\zz_{\cS^{k-1}})
 \notag
 \\*
 &\quad\cdot
 \sum_{\substack{
   \mm_{\cS_k}\notin\E((f,g)_{\cS_k},\cc_{\cS_k})
 }}
 \lrbar{
  \frac1{
   |\C_{\cS_k}|
   |\M_{\cS_k}|
  }
  -
  \mu_{Z_{\cS_k}|Z_{\cS^{k-1}}}(
   \fC_{(f,g)_{\cS_k}}(\cc_{\cS_k},\mm_{\cS_k})
   |\zz_{\cS^{k-1}}
  )
 }
 \prod_{k'=k+1}^{|\fI|}
 \frac1{
  |\C_{\cS_{k'}}|
 }
 \notag
 \\
 &=
 \sum_{k=1}^{|\fI|}
 \lrB{
  \prod_{k'=1}^k
  |\C_{\cS_{k'}}|
 }
 \sum_{
   \zz_{\cS^{k-1}}\in\fC_{f_{\cS^{k-1}}}(\cc_{\cS^{k-1}})
 }
 \mu_{Z_{\cS^{k-1}}}(\zz_{\cS^{k-1}})
 \sum_{
   \mm_{\cS_k}\in\M_{\cS_k}
 }
 \lrbar{
  \frac1{
   |\C_{\cS_k}|
   |\M_{\cS_k}|
  }
  -
  \mu_{Z_{\cS_k}|Z_{\cS^{k-1}}}(
   \fC_{(f,g)_{\cS_k}}(\cc_{\cS_k},\mm_{\cS_k})
   |\zz_{\cS^{k-1}}
  )
 }
 \notag
 \\*
 &\quad
 -
 \sum_{k=1}^{|\fI|}
 \lrB{
  \prod_{k'=1}^{k-1}
  |\C_{\cS(k')}|
 }
 \sum_{
   \zz_{\cS^{k-1}}\in\fC_{f_{\cS^{k-1}}}(\cc_{\cS^{k-1}})
 }
 \mu_{Z_{\cS^{k-1}}}(\zz_{\cS^{k-1}})
 \sum_{
  \mm_{\cS_k}\in\E((f,g)_{\cS_k},\cc_{\cS_k})
 }
 \frac1{
  |\M_{\cS_k}|
 },
 \label{eq:channel-error-2-1}
\end{align}
where the first inequality comes from
Lemma \ref{lem:diff-prod} in Appendix \ref{sec:proof-lemma}
and the fact that
$\mm_{\cS}\notin\E((f,g)_{\cS},\cc_{\cS})$
iff
$\mm_{\cS_k}\notin\E((f,g)_{\cS_k},\cc_{\cS_k})$
for all $k\in\{1,\ldots,|\fI|\}$;
the second inequality comes from the fact that
$\mm_{\cS_{k'}}\notin\E((f,g)_{\cS_{k'}},\cc_{\cS_{k'}})$ implies
\begin{align}
 \sum_{
   \zz_{\cS_{k'}}
   \in\fC_{(f,g)_{\cS_{k'}}}(\cc_{\cS_{k'}},\mm_{\cS_{k'}})
 }
 \mu_{Z_{\cS_{k'}}|Z_{\cS^{k'-1}}}(
  \zz_{\cS_{k'}}|\zz_{\cS^{k'-1}}
 )
 &=
 \mu_{Z_{\cS_{k'}}|Z_{\cS^{k'-1}}}(
  \fC_{(f,g)_{\cS_{k'}}}(\cc_{\cS_{k'}},\mm_{\cS_{k'}})
  |\zz_{\cS^{k'-1}})
 \notag
 \\
 &>0
\end{align}
for all $k'\in\{|\fI|,|\fI|-1,\ldots,k\}$
and
\begin{align}
 \sum_{\substack{
   \mm_{\cS_{k'}}\notin\E((f,g)_{\cS_{k'}},\cc_{\cS_{k'}})
   \\
   \zz_{\cS_{k'}}
   \in\fC_{(f,g)_{\cS_{k'}}}(\cc_{\cS_{k'}},\mm_{\cS_{k'}})
 }}
 \frac{
  \mu_{Z_{\cS_{k'}}|Z_{\cS^{k'-1}}}(
   \zz_{\cS_{k'}}|\zz_{\cS^{k'-1}}
  )
 }{
  \mu_{Z_{\cS_{k'}}|Z_{\cS^{k'-1}}}(
   \fC_{(f,g)_{\cS_{k'}}}(\cc_{\cS_{k'}},\mm_{\cS_{k'}})
   |\zz_{\cS^{k'-1}})
  |\M_{\cS_{k'}}|
 }
 &=
 \sum_{\substack{
   \mm_{\cS_{k'}}\notin\E((f,g)_{\cS_{k'}},\cc_{\cS_{k'}})
 }}
 \frac1{
  |\M_{\cS_{k'}}|
 }
 \notag
 \\
 &\leq
 1
\end{align}
for all $k'\in\{|\fI|,|\fI|-1,\ldots,k+1\}$.
The last equality comes from the fact that
$\mm_{\cS_k}\in\E((f,g)_{\cS_k},\cc_{\cS_k})$
implies
\begin{equation*}
 \mu_{Z_{\cS_k}|Z_{\cS^{k-1}}}(
  \fC_{(f,g)_{\cS_k}}(\cc_{\cS_k},\mm_{\cS_k})
  |\zz_{\cS^{k-1}}
 )
 =
 0.
\end{equation*}

Let $C_{\cS}\equiv\{C_s\}_{s\in\cS}$ be the set of random variables
corresponding to the uniform distribution on $\C_{\cS}$.
We have
\begin{align}
 &
 E_{(F,G)_{\cS}C_{\cS}}\lrB{
  \Error(F_{\cS},G_{\cS},C_{\cS})
 }
 \notag
 \\*
 &\leq
 E_{(F,G)_{\cS}C_{\cS}}\left[
  \sum_{k=1}^{|\fI|}
  \lrB{
   \prod_{k'=1}^k
   |\C_{\cS_{k'}}|
  }
  \sum_{
   \zz_{\cS^{k-1}}\in\fC_{F_{\cS^{k-1}}}(C_{\cS^{k-1}})
  }
  \mu_{Z_{\cS^{k-1}}}(\zz_{\cS^{k-1}})
 \right.
 \notag
 \\*
 &\qquad\cdot
 \left.
  \vphantom{
   \sum_{
    \zz_{\cS^{k-1}}\in\fC_{F_{\cS^{k-1}}}(C_{\cS^{k-1}})
   }
  }
  \sum_{
   \mm_{\cS_k}\in\M_{\cS_k}
  }
  \lrbar{
   \frac{
    1
   }{
    |\C_{\cS_k}|
    |\M_{\cS_k}|
   }
   -
   \mu_{Z_{\cS_k}|Z_{\cS^{k-1}}}(
    \fC_{(F,G)_{\cS_k}}(C_{\cS_k},\mm_{\cS_k})
    |\zz_{\cS^{k-1}}
   )
  }
 \right]
 \notag
 \\*
 &\quad
 +
 E_{(F,G)_{\cS}C_{\cS}}\lrB{
  \sum_{j\in\J}
  |\C_{\cS}|
  \sum_{\substack{
    \mm_{\cS}\in\M_{\cS},
    \zz_{\cS}\in\fC_{(F,G)_{\cS}}(C_{\cS},\mm_{\cS}),
    \\
    \yy_{\J}\in\Y^n_{\J},
    \hzz_{\D(\J)}\in\Z^n_{\D(\J)}:
    \\
    \hzz_{\D(j)}\neq \zz_{\D(j)}
  }}
  \mu_{\hZ_{\D(\J)}|C_{\cS}Y_{\J}}(\hzz_{\D(\J)}|C_{\cS},\yy_{\J})
  \mu_{Z_{\cS}\YY_{\J}}(\zz_{\cS},\yy_{\J})
 }
 \notag
 \\
 &=
 \sum_{k=1}^{|\fI|}
 E_{(F,G)_{\cS_k}}\left[
  \sum_{
   \zz_{\cS^{k-1}}\in\Z^n_{\cS^{k-1}}
  }
  \mu_{Z_{\cS^{k-1}}}(\zz_{\cS^{k-1}})
  \sum_{
   \cc_{\cS_k}\in\C_{\cS_k},
   \mm_{\cS_k}\in\M_{\cS_k}
  }
  \lrbar{
   \frac{
    1
   }{
    |\C_{\cS_k}|
    |\M_{\cS_k}|
   }
   -
   \mu_{Z_{\cS_k}|Z_{\cS^{k-1}}}(
    \fC_{(F,G)_{\cS_k}}(\cc_{\cS_k},\mm_{\cS_k})
    |\zz_{\cS^{k-1}}
   )
  }
  \right]
 \notag
 \\*
 &\quad
 +
 \sum_{j\in\J}
  E_{F_{\D(j)}}\lrB{
   \sum_{\substack{
    \cc_{\D(j)}\in\C_{\D(j)},
    \zz_{\D(j)}\in\fC_{F_{\D(j)}}(\cc_{\D(j)}),
    \yy_j\in\Y^n_j,
    \hzz_{\D(j)}\in\Z^n_{\D(j)}:
    \\
    \hzz_{\D(j)}\neq \zz_{\D(j)}
  }}
  \mu_{\hZ_{\D(j)}|Y_jC_{\D(j)}}
  (\hzz_{\D(j)}|\yy_j,\cc_{\D(j)})
  \mu_{Z_{\D(j)}Y_j}(\zz_{\D(j)},\yy_j)
 }
 \notag
 \\
 &=
 \sum_{k=1}^{|\fI|}
 E_{(F,G)_{\cS_k}}\lrB{
  \sum_{\substack{
   \zz_{\ocS_k}\in\Z^n_{\ocS_k},
   \cc_{\cS_k}\in\C_{\cS_k},
   \mm_{\cS_k}\in\M_{\cS_k}
  }}
  \mu_{Z_{\ocS_k}}(\zz_{\ocS_k})
  \lrbar{
   \frac{
    1
   }{
    |\C_{\cS_k}|
    |\M_{\cS_k}|
   }
   -
   \mu_{Z_{\cS_k}|Z_{\ocS_k}}(
    \fC_{(F,G)_{\cS_k}}(\cc_{\cS_k},\mm_{\cS_k})
    |\zz_{\ocS_k})
  }
 }
 \notag
 \\*
 &\quad
 +
 \sum_{j\in\J}
 E_{F_{\D(j)}}\lrB{
  \mu_{Z_{\D(j)}\hZ_{\D(j)}}\lrsb{\lrb{
    (\zz_{\D(j)},\hzz_{\D(j)}):
    \hzz_{\D(j)}\neq\zz_{\D(j)}
   }
  }
 }
 \notag
 \\
 &\leq
 \sum_{k=1}^{|\fI|}
 \sqrt{\textstyle
  \alpha_{(F,G)_{\cS_k}}-1
  +\sum_{
    \cS'\subset\cS_k:
    \cS'\neq\emptyset
  }
  \alpha_{(F,G)_{\cS_k\setminus\cS'}}
  [\beta_{(F,G)_{\cS'}}+1]
  2^{-n\gamma(k,\cS')}
 }
 +
 2
 \sum_{k=1}^{|\fI|}
 \mu_{Z_{\ocS_k\cup\cS_k}}(\uT_k^{\complement})
 \notag
 \\*
 &\quad
 +2\sum_{j\in\J}
  \sum_{\substack{
    \D'\subset\D(j):
    \D'\neq\emptyset
  }}
  \alpha_{F_{\D'}}\lrB{\beta_{F_{\D(j)\setminus\D'}}+1}
  2^{
   -n\gamma(j,\D')
  }
  +2 \sum_{j\in\J}
  \beta_{F_{\D(j)}}
  +2 \sum_{j\in\J}
  \mu_{Z_{\D(j)}Y_j}(\oT_j^{\complement}),
 \label{eq:proof-channel-error-ave}
\end{align}
where
\begin{align*}
 \gamma(k,\cS')
 &\equiv
 \uH(\ZZ_{\cS'}|\ZZ_{\ocS_k})-\sum_{s\in\cS'}[r_s+R_s]-\e
 \\
 \gamma(j,\D')
 &\equiv
 \sum_{s\in\D'}r_s-\oH(\ZZ_{\D'}|\YY_j,\ZZ_{\D(j)\setminus\D'})-\e
 \\
 \uT_k
 &\equiv
 \lrb{
  (\zz_{\cS_k},\zz_{\ocS_k}):
  \begin{aligned}
   &\frac 1n
   \log_2\frac1{
    \mu_{Z^n_{\cS'}|Z^n_{\ocS_k}}(\zz_{\cS'}|\zz_{\ocS_k})
   }
   \geq
   \uH(\ZZ_{\cS'}|\ZZ_{\ocS_k})-\e
   \\
   &\text{for all}\ \cS'\subset\cS_k
   \ \text{satisfying}\ \emptyset\neq\cS'\subset\cS_k
  \end{aligned}
 }
 \\
 \oT_j
 &\equiv
 \lrb{
  (\zz_{\D(j)},\yy):
  \begin{aligned}
   &
   \frac1n\log
   \frac1{
    \mu_{Z_{\D'}|Z_{\D(j)\setminus\D'}Y_j}
    (\zz_{\D'}|\zz_{\D(j)\setminus\D'},\yy_j)
   }
   \leq
   \oH(\ZZ_{\D'}|\YY_j,\ZZ_{\D(j)\setminus\D'})+\e
   \\
   &\text{for all}\ \D'\ \text{satisfying}\ \emptyset\neq\D'\subset\D(j)
  \end{aligned}
 }.
\end{align*}
The first inequality comes from
(\ref{eq:channel-error})--(\ref{eq:channel-error-2-1})
and the fact that
\begin{align}
 E_{C_{\cS^{k-1}}}\lrB{
  \lrB{
   \prod_{k'=1}^{k-1}
   |\C_{\cS_{k'}}|
  }
  \sum_{
   \zz_{\cS^{k-1}}\in\fC_{f_{\cS^{k-1}}}(\cc_{\cS^{k-1}})
  }
  \mu_{Z_{\cS^{k-1}}}(\zz_{\cS^{k-1}})
 } 
 &=
 \sum_{
  \cc_{\cS^{k-1}}\in\C_{\cS^{k-1}},
  \zz_{\cS^{k-1}}\in\fC_{f_{\cS^{k-1}}}(\cc_{\cS^{k-1}})
 }
 \mu_{Z_{\cS^{k-1}}}(\zz_{\cS^{k-1}})
 \notag
 \\
 &=
 1
 \label{eq:proof-channel-ave-CS(1:k-1)}
\end{align}
implies
\begin{align*}
 &
 E_{F_{\cS}G_{\cS}C_{\cS}}\lrB{
 \sum_{k=1}^{|\fI|}
 \lrB{
  \prod_{k'=1}^{k-1}
  |\C_{\cS_{k'}}|
 }
 \sum_{
  \zz_{\cS^{k-1}}\in\fC_{f_{\cS^{k-1}}}(\cc_{\cS^{k-1}})
 }
 \mu_{Z_{\cS^{k-1}}}(\zz_{\cS^{k-1}})
 \sum_{
  \mm_{\cS_k}\in\E((F,G)_{\cS_k},C_{\cS_k})
 }
 \frac1{
  |\M_{\cS_k}|
 }
 }
 \notag
 \\*
 &=
 \sum_{k=1}^{|\fI|}
 E_{F_{\cS_k}G_{\cS_k}C_{\cS_k}}\lrB{
  \sum_{
   \mm_{\cS_k}\in\E((F,G)_{\cS_k},C_{\cS_k})
  }
  \frac{
   1
  }{
   |\M_{\cS_k}|
  }
 },
\end{align*}
the second equality comes from (\ref{eq:mu_ZSk|ZoSk}),
and the last inequality comes from
Lemmas~\ref{lem:hash-FG}, \ref{lem:channel-bcp}, \ref{lem:channel-crp}
in Appendixes \ref{sec:hash}--\ref{sec:channel-crp},
and the relations $r_s=\log_2(|\C_s|)=\log_2(|\im\F_s|)/n$,
$R_s=\log_2(|\M_s|)/n=\log_2(|\im\G_s|)/n$.

Finally, let us assume that $\{(r_s,R_s)\}_{s\in\cS}$ satisfies
(\ref{eq:rate-positive})--(\ref{eq:rate-decoder})
for all $(k,\cS')$ satisfying
$k\in\{1,\ldots,|\fI|\}$, $\emptyset\neq\cS'\subset\cS_k$,
and $(j,\D')$ satisfying $j\in\J$, $\emptyset\neq\D'\subset\D(j)$.
We have $\gamma(k,\cS')>0$ and $\gamma(j,\D')>0$ for all $(k,\cS',j,\D')$
satisfying $k\in\{1,\ldots,|\fI|\}$, $\emptyset\neq\cS'\subset\cS_k$,
$j\in\J$, $\emptyset\neq\D'\subset\D(j)$.
Then, by letting
$\alpha_{F_s}\to1$, $\beta_{F_s}\to0$,
$\alpha_{G_s}\to1$, $\beta_{G_s}\to0$,
$\mu_{Z_{\ocS_k\cup\cS_k}}(\uT_k^{\complement})\to0$,
$\mu_{Z_{\D(j)}Y_j}(\oT_j^{\complement})\to0$,
$\e\to0$,
and using the random coding argument,
we have the fact that
for all $\delta>0$ and sufficiently large $n$
there are $f_{\cS}=\{f_s\}_{s\in\cS}$, $g_{\cS}=\{g_s\}_{s\in\cS}$,
and $\cc_{\cS}=\{\cc_s\}_{s\in\cS}$
such that $\Error(f_{\cS},g_{\cS},\cc_{\cS})\leq\delta$.
\hfill\IEEEQED
  
\appendix
\subsection{Algorithm for Linear Extension of
 Reversed Partial Ordering of Subsets}
\label{sec:tsort}

This section introduces
an algorithm for computing the linear extension of the reversed partial ordering
of $\fI\equiv\{\I_1,\ldots,\I_{|\fI|}\}\subset 2^{\I}$
which yields the following property:
$\I_k\subsetneq\I_{k'}$ implies $k'<k$
for all $k,k'\in\{1,2,\ldots,|\fI|\}$.

When a partial ordering is represented by a directed acyclic graph,
the standard topological sort \cite[Section 22.4]{CLRS09}
can be employed with time complexity of $O(|\fI|+a)$,
where $a$ is the number of arcs.
However, 
when no directed acyclic graph is available,
the construction has time complexity of $O(|\I||\fI|^2)$,
where factor $O(|\I|)$ corresponds
to the computation of partial ordering of two subsets
and $O(|\fI|^2)$ corresponds to the combination of two vertexes.
We may also use a variation of the quick sort \cite{DKMRV07}
which could be employed with expected time complexity of
$O(w|\I||\fI|\log_2|\fI|)$,
where $w$ is the width of the directed acyclic graph
and factor $O(|\I|)$ corresponds
to the computation of partial ordering of two subsets.

Here, let us introduce Algorithm \ref{alg:tsort}
similar to the bucket sort \cite[Section 8.4]{CLRS09},
based on the cardinality of a subset.
When the time complexity of computing the cardinality of a subset
is $O(1)$, this algorithm has time complexity of $O(|\fI|+|\I|)$.
When the time complexity of computing the cardinality of a subset is $O(|\I|)$,
this algorithm has time complexity of $O(|\I||\fI|)$.
It is assumed that each $\fI$ and $\fL(v)$, $v\in\{0,1,\ldots,|\I|\}$
is represented as a list of subsets.
Line~1 corresponds to the initialization of $v\in\{0,1,\ldots,|\I|\}$.
At Line~2, $\fI\leftarrow\{\I_k\}\cup\fL(|\I_k|)$ means that
$\{\I_k\}$ is appended to $\fL(|\I_k|)$.
At Line~4, $\fI\leftarrow\fL(v)\cup\fI$ means that
$\fL(v)$ is appended to the beginning of $\fI$.
We can implement $\fI$ and $\fL(w)$, $w\in\{0,1,\ldots,|\I|\}$
by using a linked list~\cite[Section 10.2]{CLRS09}
to reduce the time complexity of the union operation.
We have the following lemma.
\begin{lem}
After employing Algorithm \ref{alg:tsort},
$\fI\equiv\{\I_1,\ldots,\I_{|\fI|}\}$ satisfies the property that
$\I_k\subsetneq\I_{k'}$ implies $k'<k$ for all $k,k'\in\{1,2,\ldots,|\fI|\}$.
\end{lem}
\begin{IEEEproof}
 After employing Algorithm \ref{alg:tsort},
 $\fI\equiv\{\I_1,\ldots,\I_{|\fI|}\}$ satisfies the property that
 $k\leq k'$ implies $|\I_{k'}|\leq|\I_k|$ for all $k,k'\in\{1,2,\ldots,|\fI|\}$.
 This is equivalent to the fact that
 $|\I_k|<|\I_{k'}|$ implies $k'<k$ for all $k,k'\in\{1,2,\ldots,|\fI|\}$.
 The above yields the fact that
 $\I_k\subsetneq\I_{k'}$ implies 
 $|\I_k|<|\I_{k'}|$ and $k'<k$ for all $k,k'\in\{1,2,\ldots,|\fI|\}$.
\end{IEEEproof}

\begin{rem}
We can also use standard sorting algorithms
(e.g. quick sort, merge sort, heap sort) based on subset cardinality; 
the time complexity is $O(|\fI|\log|\fI|)$
when the time complexity for computing the cardinality of subset $\I'\in\fI$ is $O(1)$.
That is, the standard sorting algorithms
could be better than the proposed algorithm
if $|\I|=\Omega(|\fI|\log|\fI|)$
(including the case when $|\I|$ is unknown/infinite).
\end{rem}

\begin{algorithm}[t]
 \caption{Linear extension of $\fI\subset 2^{\I}$}
 \hspace*{\algorithmicindent}\textbf{Input:}
 List $\fI\equiv\{\I_1,\ldots,\I_{|\fI|}\}$.
 \\
 \hspace*{\algorithmicindent}\textbf{Output:}
 Linear extension of $\fI$,
 where $\I_k\subsetneq\I_{k'}$
 implies $k'<k$ for all $k,k'\in\{1,2,\ldots,|\fI|\}$.
 \label{alg:tsort}
 \begin{algorithmic}[1]
  \FFor{$v\in\{0,\ldots,|\I|\}$} $\fL(v)\leftarrow\emptyset$.
  \FFor{$k\in\{1,\ldots,|\fI|\}$}
  $\fL(|\I_k|)\leftarrow\{\I_k\}\cup\fL(|\I_k|)$.
  \State $\fI\leftarrow\emptyset$.
  \FFor{$v\in\{0,\ldots,|\I|\}$} $\fI\leftarrow\fL(v)\cup\fI$.
 \end{algorithmic}
\end{algorithm}

\subsection{Information-Spectrum Methods}
\label{sec:ispec}

First, we review the definition of the limit superior/inferior in
probability introduced in \cite{HAN}.
For sequence $\{U_n\}_{n=1}^{\infty}$ of random variables,
the {\em limit superior in probability} $\plimsupn U_n$
and the {\em limit inferior in probability} $\pliminfn U_n$ are defined
as
\begin{align*}
 \plimsupn U_n
 &\equiv 
 \inf\lrb{\theta: \limn \Prob\lrsb{U_n>\theta}=0}
 \\
 \pliminfn U_n
 &\equiv
 \sup\lrb{\theta: \limn \Prob\lrsb{U_n<\theta}=0}.
\end{align*}
Moreover, we have the following relations~\cite[Section 1.3]{HAN}:
\begin{align}
 \pliminfn\lrB{U_n+V_n}
 &\leq \plimsupn U_n + \pliminfn V_n
 \label{eq:pliminf-upper}
 \\
 \plimsupn\lrB{-U_n}
 &=\pliminfn U_n.
 \label{eq:plimsup-pliminf}
\end{align}

For sequence $\{\mu_{U_n}\}_{n=1}^{\infty}$
of probability distributions corresponding to $\UU$,
we define the spectral inf-entropy rate, $\uH(\UU)$, as
\begin{align*}
 \uH(\UU)
 &\equiv
 \pliminfn\frac 1n\log_2\frac1{\mu_{U_n}(U_n)}.
\end{align*}
For general sequence $\{\mu_{U_nV_n}\}_{n=1}^{\infty}$ of
joint probability distributions
corresponding to $(\UU,\VV)=\{(U_n,V_n)\}_{n=1}^{\infty}$,
we define the spectral conditional sup-entropy rate $\oH(\UU|\VV)$,
and the spectral conditional inf-entropy rate $\uH(\UU|\VV)$ as
\begin{align*}
 \oH(\UU|\VV)
 &\equiv
 \plimsupn\frac 1n\log_2\frac1{\mu_{U_n|V_n}(U_n|V_n)}
 \\
 \uH(\UU|\VV)
 &\equiv
 \pliminfn\frac 1n\log_2\frac1{\mu_{U_n|V_n}(U_n|V_n)}.
\end{align*}

In the following, we introduce some inequalities that we use in the
proof of the converse part.
Trivially, we have
\begin{equation*}
 \oH(\UU|\VV)\geq\uH(\UU|\VV)\geq 0.
\end{equation*}

We show the following lemmas.
\begin{lem}[{\cite[Lemma 3.2.1, Definition 4.1.3]{HAN}}]
\label{lem:pliminf-div}
For general sources $\UU\equiv\{U_n\}_{n=1}^{\infty}$ and
$\VV\equiv\{V_n\}_{n=1}^{\infty}$, we have
\begin{equation*}
 \pliminfn\frac 1n\log_2\frac{\mu_{U_n}(U_n)}{\mu_{V_n}(U_n)}\geq 0.
\end{equation*}
\end{lem}
\begin{IEEEproof}
For completeness,
we show the lemma as the proof of \cite[Lemma 3.2.1, Definition 4.1.3]{HAN}.
For a given $\gamma>0$, let $\U'_n$ be defined as
\begin{equation}
 \U'_n
 \equiv\lrb{
  \uu: \frac 1n\log_2\frac{\mu_{U_n}(\uu)}{\mu_{V_n}(\uu)}
  < -\gamma
 }.
\end{equation}
Then we have $\mu_{U_n}(\uu)<\mu_{V_n}(\uu)2^{-n\gamma}$ for all $\uu\in\U'_n$.
We have
\begin{align}
 \Prob\lrsb{
  \frac 1n\log_2\frac{\mu_{U_n}(U_n)}{\mu_{V_n}(U_n)}< -\gamma
 }
 &=
 \sum_{\uu\in\U'_n}
 \mu_{U_n}(\uu)
 \notag
 \\
 &<
 \sum_{\uu\in\U'_n}
 \mu_{V_n}(\uu)2^{-n\gamma}
 \notag
 \\
 &\leq
 2^{-n\gamma},
\end{align}
which implies
\begin{equation*}
 \pliminfn
 \frac 1n\log_2\frac{\mu_{U_n}(\uu)}{\mu_{V_n}(\uu)}\geq -\gamma
\end{equation*}
from the definition of the limit inferior in probability.
The  lemma is proven by letting $\gamma\to0$.
\end{IEEEproof}

\begin{lem}
\label{lem:oH(U|V)>oH(U|VW)}
For a triplet of general sources
$(\UU,\VV,\WW)=\{(U_n,V_n,W_n)\}_{n=1}^{\infty}$,
we have
$\oH(\UU|\VV)\geq \oH(\UU|\VV,\WW)$.
\end{lem}
\begin{IEEEproof}
 We have
 \begin{align}
  \oH(\UU|\VV)-\oH(\UU|\VV,\WW)
  &=
  \plimsupn\frac 1n\log_2
  \frac1{\mu_{U_n|V_n}(U_n|V_n)}
  -
  \plimsupn\frac 1n\log_2\frac1{\mu_{U_n|V_nW_n}(U_n|V_n,W_n)}
  \notag
  \\*
  &=
  \plimsupn\frac 1n\log_2
  \frac1{\mu_{U_n|V_n}(U_n|V_n)}
  +
  \pliminfn\frac 1n\log_2\mu_{U_n|V_nW_n}(U_n|V_n,W_n)
  \notag
  \\
  &\geq
  \pliminfn\frac 1n\log_2
  \frac{\mu_{U_n|V_nW_n}(U_n|V_n,W_n)}{\mu_{U_n|V_n}(U_n|V_n)}
  \notag
  \\
  &=
  \pliminfn\frac 1n\log_2
  \frac{\mu_{U_nV_nW_n}(U_n,V_n,W_n)}
  {\mu_{U_n|V_n}(U_n|V_n)\mu_{V_nW_n}(V_n,W_n)}
  \notag
  \\
  &\geq
  0,
 \end{align}
 where the second equality comes from (\ref{eq:plimsup-pliminf}),
 the first inequality comes from (\ref{eq:pliminf-upper}),
 and the second inequality comes from Lemma~\ref{lem:pliminf-div}.
\end{IEEEproof}

The following lemma is analogous to the Fano inequality.
\begin{lem}[{\cite[Lemma 4]{K08}\cite[Lemma 7]{CRNG}}]
\label{lem:fano}
Let $(\UU,\VV)\equiv\{(U_n,V_n)\}_{n=1}^{\infty}$ be a sequence
of two random variables.
If there is a sequence $\{\Psi_n\}_{n=1}^{\infty}$
of (possibly stochastic) functions independent of $(\UU,\VV)$
that satisfy the condition
\begin{equation}
 \limn \Prob(\Psi_n(V_n)\neq U_n)=0,
 \label{eq:fano-error}
\end{equation}
then
\begin{equation*}
 \oH(\UU|\VV)=0.
\end{equation*}
\end{lem}
\begin{IEEEproof}
When $\{\Psi_n\}_{n=1}^{\infty}$ is a sequence of deterministic
functions, the lemma is the same as \cite[Lemma 7]{CRNG}.
For completeness,
we show this lemma following to the proof of \cite[Lemma 1.3.2]{HAN}.

Let $\{\psi_n\}_{n=1}^{\infty}$ be a sequence of deterministic
functions satisfying
\begin{equation}
 \limn P(\psi_n(V_n)\neq U_n)=0.
 \label{eq:fano-error-deterministic}
\end{equation}
For $\gamma>0$, let
\begin{align*}
  \W
  &\equiv\lrb{
    (\uu,\vv): \frac 1n\log_2\frac1{\mu_{U_n|V_n}(\uu|\vv)}\geq\gamma
  }
  \\
  \E
  &\equiv\lrb{
    (\uu,\vv): \psi_n(\vv)\neq\uu
  }.
\end{align*}
Then we have
\begin{align}
  \Prob\lrsb{ \frac 1n\log_2\frac1{\mu_{U_n|V_n}(U_n|V_n)} > \gamma}
  &\leq
  \mu_{U_nV_n}(\W)
  \notag
  \\
  &=
  \mu_{U_nV_n}(\W\cap\E)
  +\mu_{U_nV_n}(\W\cap\E^{\complement})
  \notag
  \\
  &=
  \mu_{U_nV_n}(\W\cap\E)
  +\sum_{(\uu,\vv)\in\W\cap\E^{\complement}}\mu_{U_nV_n}(\uu,\vv)
  \notag
  \\
  &=
  \mu_{U_nV_n}(\W\cap\E)
  +
  \sum_{\vv\in\V_n}\mu_{V_n}(\vv)
  \sum_{\substack{
    \uu\in\U_n:\\
    \psi_n(\vv)=\uu\\
    (\uu,\vv)\in\W
  }}
  \mu_{U_n|V_n}(\uu|\vv)
  \notag
  \\
  &\leq
  \mu_{U_nV_n}(\W\cap\E)
  +
  \sum_{\vv\in\V_n}\mu_{V_n}(\vv)
  \sum_{\uu\in\U_n: \psi_n(\vv)=\uu}
  2^{-n\gamma}
  \notag
  \\
  &\leq
  P(\psi_n(V_n)\neq U_n)+2^{-n\gamma},
\end{align}
where the second inequality comes from the definition of $\W$
and the last inequality comes from the fact that
for all $\vv$ there is a unique $\uu$ satisfying $\psi_n(\vv)=\uu$.
From this inequality and (\ref{eq:fano-error}), we have
\[
  \limn\Prob\lrsb{
    \frac 1n\log_2\frac1{\mu_{U_n|V_n}(U_n|V_n)}>\gamma
  }=0.
\]
Then we have
\[
  0\leq\oH(\UU|\VV)\leq \gamma
\]
from the definition of $\oH(\UU|\VV)$.
We have $\oH(\UU|\VV)=0$ by letting $\gamma\to0$.

When $\{\Psi_n\}_{n=1}^{\infty}$ is a sequence of stochastic functions,
we can obtain sequence $\{\psi_n\}_{n=1}^{\infty}$ of deterministic functions
such that
\begin{align*}
 \Prob(\psi_n(V_n)\neq U_n)
 &\leq
 \sum_{\psi_n}\Prob(\Psi_n=\psi_n)\Prob(\psi_n(V_n)\neq U_n)
 \\
 &=
 \Prob(\Psi_n(V_n)\neq U_n)
\end{align*}
for all $n\in\NN$ from the random coding argument
and the fact that $\Psi_n$ is independent of $(U_n,V_n)$.
Then we have the fact that
(\ref{eq:fano-error}) implies (\ref{eq:fano-error-deterministic})
and $\oH(\UU|\VV)=0$.
\end{IEEEproof}

\subsection{$(\aalpha,\bbeta)$-hash property}
\label{sec:hash}

In this section, we review the hash property
introduced in \cite{CRNG}\cite{HASH-BC} and show two basic lemmas.
For set $\F$ of functions,
let $\im\F\equiv \bigcup_{f\in\F}\{f(\zz): \zz\in\Z^n\}$.

\begin{df}[{\cite[Definition~3]{CRNG}}]
Let $\F_n$ be a set of functions on $\U^n$.
For probability distribution $p_{F_n}$ on $\F_n$, we
call pair $(\F_n,p_{F_n})$ an {\em ensemble}.
Then, $(\F_n,p_{F_n})$ has an $(\alpha_{F_n},\beta_{F_n})$-{\em hash property} if
there is pair $(\alpha_{F_n},\beta_{F_n})$
depending on $p_{F_n}$ such that
\begin{align}
 \sum_{\substack{
   \zz'\in\Z^n\setminus\{\zz\}:
   \\
   p_{F_n}(\{f: f(\zz) = f(\zz')\})>\frac{\alpha_{F_n}}{|\im\F_n|}
 }}
 p_{F_n}\lrsb{\lrb{f: f(\zz) = f(\zz')}}
 \leq
 \beta_{F_n}
 \label{eq:hash}
\end{align}
for any $\zz\in\Z^n$.
Consider the following conditions for two sequences
$\aalpha_F\equiv\{\alpha_{F_n}\}_{n=1}^{\infty}$ and
$\bbeta_F\equiv\{\beta_{F_n}\}_{n=1}^{\infty}$
\begin{align}
 \limn \alpha_{F_n}
 &=1
 \label{eq:alpha-bcp}
 \\
 \limn \beta_{F_n}
 &=0.
 \label{eq:beta-crp}
\end{align}
Then, we can say that 
$(\bcF,\bp_F)$ has an $(\aalpha_F,\bbeta_F)$-{\em hash property}
if $\aalpha_F$ and $\bbeta_F$
satisfy (\ref{eq:hash})--(\ref{eq:beta-crp}).
Throughout this paper,
we omit the dependence of $\F$ and $F$ on $n$.
\end{df}

It should be noted that
when $\F$ is a two-universal class of hash functions \cite{CW79}
and  $p_F$ is the uniform distribution on $\F$,
then $(\bcF,\bp_F)$ has a $(\one,\zero)$-hash property,
where $\one\equiv(1,1,\ldots)$ and $\zero\equiv(0,0,\ldots)$.
Random binning \cite{C75} and the set of all linear functions \cite{CSI82}
are two-universal classes of hash functions.
It is proved in \cite[Section III-B]{HASH-BC} that
an ensemble of sparse matrices has a hash property.

First, we introduce the lemma for a joint ensemble.
\begin{lem}[
 {\cite[Lemma 4 of the extended version]{HASH-BC}\cite[Lemma 3]{CRNG}}
]
\label{lem:hash-FG}
Let $(\F,p_F)$ and $(\G,p_G)$ be ensembles of functions on the same set $\Z^n$.
Assume that $(\F,p_F)$ (resp. $(\G,p_G)$) has an $(\alpha_F,\beta_F)$-hash
(resp. $(\alpha_G,\beta_G)$-hash) property.
Let $(f,g)\in\F\times\G$ be a function defined as
\begin{equation*}
 (f,g)(\zz)\equiv(f(\zz),g(\zz))\quad\text{for each}\ \zz\in\Z^n.
\end{equation*}
Let $p_{(F,G)}$  be a joint distribution on $\F\times\G$ defined as
\begin{equation*}
 p_{(F,G)}(f,g)\equiv p_F(f)p_G(g)\quad\text{for each}\ (f,g)\in\F\times\G.
\end{equation*}
Then the ensemble $(\F\times\G, p_{(F,G)})$ has an
$(\alpha_{(F,G)},\beta_{(F,G)})$-hash property,
where $\alpha_{(F,G)}$ and $\beta_{(F,G)}$ are defined as
\begin{align*}
 \alpha_{(F,G)}
 &\equiv
 \alpha_F\alpha_G
 \\
 \beta_{(F,G)}
 &\equiv
 \beta_F+\beta_G.
\end{align*}
\end{lem}
\begin{IEEEproof}
We show this lemma for completeness.
Let
\begin{align*}
 p_{F,\zz,\zz'}
 &
 \equiv
 p_F(\{f: f(\zz)=f(\zz')\})
 \\
 p_{G,\zz,\zz'}
 &\equiv
 p_G(\{g: g(\zz)=g(\zz')\})
 \\
 p_{(F,G),\zz,\zz'}
 &\equiv
 p_{(F,G)}(
  \{(f,g): (f,g)(\zz)=(f,g)(\zz')\}
 ).
\end{align*}
Then we have
\begin{align}
 &
 \sum_{\substack{
   \zz'\in\Z^n\setminus\{\zz\}:
   \\
   p_{(F,G),\uu,\uu'}>\frac{\alpha_{(F,G)}}{|\im\F\times\G|}
 }}
 p_{(F,G)}(\{(f,g): (f,g)\zz=(f,g)\zz'\})
 \notag
 \\*
 &\leq
 \sum_{\substack{
   \zz'\in\Z^n\setminus\{\zz\}:
   \\
   p_{F,\zz,\zz'}p_{G,\zz,\zz'}>\frac{\alpha_F\alpha_G}{|\im\F||\im\G|}
 }}
 p_{F,\zz,\zz'}p_{G,\zz,\zz'}
 \notag
 \\
 &=
 \sum_{\substack{
   \zz'\in\Z^n\setminus\{\zz\}:
   \\
   p_{F,\zz,\zz'}p_{G,\zz,\zz'}>\frac{\alpha_F\alpha_G}{|\im\F||\im\G|}
   \\
   p_{F,\zz,\zz'}>\frac{\alpha_F}{|\im\F|}
 }}
 p_{F,\zz,\zz'}p_{G,\zz,\zz'}
 +
 \sum_{\substack{
   \zz'\in\Z^n\setminus\{\zz\}:
   \\
   p_{F,\zz,\zz'}p_{G,\zz,\zz'}>\frac{\alpha_F\alpha_G}{|\im\F||\im\G|}
   \\
   p_{F,\zz,\zz'}\leq\frac{\alpha_F}{|\im\F|}
 }}
 p_{F,\zz,\zz'}p_{G,\zz,\zz'}
 \notag
 \\
 &\leq
 \sum_{\substack{
   \zz'\in\Z^n\setminus\{\zz\}:
   \\
   p_{F,\zz,\zz'}>\frac{\alpha_F}{|\im\F|}
 }}
 p_{F,\zz,\zz'}p_{G,\zz,\zz'}
 +
 \sum_{\substack{
   \zz'\in\Z^n\setminus\{\zz\}:
   \\
   p_{G,\zz,\zz'}>\frac{\alpha_G}{|\im\G|}
 }}
 p_{F,\zz,\zz'}p_{G,\zz,\zz'}
 \notag
 \\
 &\leq
 \sum_{\substack{
   \zz'\in\Z^n\setminus\{\zz\}:
   \\
   p_{F,\zz,\zz'}>\frac{\alpha_F}{|\im\F|}
 }}
 p_{F,\zz,\zz'}
 +
 \sum_{\substack{
   \zz'\in\Z^n\setminus\{\zz\}:
   \\
   p_{G,\zz,\zz'}>\frac{\alpha_G}{|\im\G|}
 }}
 p_{G,\zz,\zz}
 \notag
 \\
 &=
 \beta_F+\beta_G
 \notag
 \\
 &=
 \beta_{(F,G)},
\end{align}
where the first inequality comes from the fact that
$F$ and $G$ are mutually independent
and $\im\F\times\G\subset\im\F\times\im\G$,
and the last inequality comes from the fact that
$p_{F,\zz,\zz'}\leq 1$ and $p_{G,\zz,\zz'}\leq 1$.
Then we have the fact that
$(\F\times\G,p_{(F,G)})$ has an
$(\alpha_{(F,G)},\beta_{(F,G)})$-hash property.
\end{IEEEproof}

Next, we introduce lemmas that are multiple extensions of
the {\it balanced-coloring property} and the {\it collision-resistant property}.
We use the following notations.
For each $s\in\cS$, let $\F_s$ be a set
of functions on $\Z_s^n$ and $\cc_s\in\im\F_s$.
Let $\Z_{\cS'}^n\equiv\Prod_{s\in\cS'}\Z_s^n$ and 
\begin{align*}
 \alpha_{F_{\cS'}}
 &\equiv
 \prod_{s\in\cS'}\alpha_{F_s}
 \\
 \beta_{F_{\cS'}}
 &\equiv
 \prod_{s\in\cS'}\lrB{\beta_{F_s}+1}-1,
\end{align*}
where $\prod_{s\in\emptyset}\theta_s\equiv1$.
It should be noted that
\begin{align*}
 \limn \alpha_{F_{\cS'}}=1
 \\
 \limn \beta_{F_{\cS'}}=0
\end{align*}
for every $\cS'\subset\cS$
when $(\aalpha_{F_s},\bbeta_{F_s})$ satisfies
(\ref{eq:alpha-bcp}) and (\ref{eq:beta-crp}) for all $s\in\cS$.
For $\T\subset\Z_{\cS}^n$ and $\zz_{\cS'}\in\Z^n_{\cS'}$,
let $\T_{\cS'}$ and $\T_{\cS'^{\complement}|\cS'}(\zz_{\cS'})$ be defined as
\begin{align*}
 &
 \T_{\cS'}
 \equiv\{\zz_{\cS'}:
  (\zz_{\cS'},\zz_{\cS'^{\complement}})\in\T
  \ \text{for some}\ \zz_{\cS'^{\complement}}\in\Z_{\cS'^{\complement}}
  \}
 \\
 &
 \T_{\cS'^{\complement}|\cS'}(\zz_{\cS'})
 \equiv
 \{\zz_{\cS'^{\complement}}: (\zz_{\cS'},\zz_{\cS'^{\complement}})\in\T\}.
\end{align*}

The following lemma is related to the {\em balanced-coloring property},
which is an extension of \cite[Lemma 4]{HASH-WTC},
the leftover hash lemma~\cite{IZ89}
and the balanced-coloring lemma~\cite[Lemma 3.1]{AC98}\cite[Lemma 17.3]{CK11}.
This lemma implies that there is an assignment that divides a set equally.
\begin{lem}[{\cite[Lemma 4 in the extended version]{CRNG-MULTI}}]
\label{lem:mBCP}
For each $s\in\cS$, let $\F_s$ be a set
of functions on $\Z_s^n$
and $p_{F_s}$ be the probability distribution on $\F_s$,
where $(\F_s,p_{F_s})$ satisfies (\ref{eq:hash}).
We assume that random variables $\{F_s\}_{s\in\cS}$ are mutually independent.
Then
\begin{align*}
 E_{F_{\cS}}\lrB{
  \sum_{\cc_{\cS}}
  \lrbar{
   \frac{Q(\T\cap\fC_{F_{\cS}}(\cc_{\cS}))}
   {Q(\T)}
   -
   \frac1
   {\prod_{s\in\cS}|\im\F_s|}
  }
 }
 &\leq
 \sqrt{
  \alpha_{F_{\cS}}-1
  +
  \!\!\!\!\!\!\!
   \sum_{\substack{
     \cS'\subset\cS:
     \cS'\neq\emptyset
   }}
  \!\!\!\!\!\!\!
   \alpha_{F_{\cS'^{\complement}}}
   \lrB{\beta_{F_{\cS}}+1}
   \lrB{\prod_{s\in\cS'}|\im\F_s|}
   \cdot
  \frac{
   \oQ_{\cS'^{\complement}}
  }
  {Q(\T)}
 }
\end{align*}
for any function $Q:\Z_{\cS}\to[0,\infty)$ and $\T\subset\Z_{\cS}^n$,
where
\begin{equation}
 \oQ_{\cS'^{\complement}}
 \equiv
 \begin{cases}
  \displaystyle
  \max_{\zz_{\cS}\in\T}Q(\zz_{\cS})
  &\!\!\text{if}\ \cS'^{\complement}=\cS
  \\
  \displaystyle
  \max_{\zz_{\cS'}\in\T_{\cS'}}
  \!\!\!
  \sum_{\zz_{\cS'^{\complement}}\in\T_{\cS'^{\complement}|\cS'}(\zz_{\cS'})}
  \!\!\!
  Q(\zz_{\cS'},\zz_{\cS'^{\complement}})
  &\!\!\text{if}\ \emptyset\neq\cS'^{\complement}\subsetneq\cS
 \end{cases}
 \label{eq:maxQJ}
\end{equation}
\end{lem}
\begin{IEEEproof}
Let
$p_{\zz_s,\zz'_s}
\equiv
p_{F_s}\lrsb{\lrb{
  f_s:
  f_s(\zz_s)=f_s(\zz'_s)
}}$
and let $C_{\cS}$ be the random variable
corresponding to the uniform distribution on $\Prod_{s\in\cS}\im\F_s$.
In the following, we use the relation
\begin{align}
 \sum_{\substack{
   \zz_s\in\Z^n_s:
   \\
   p_{\zz_s,\zz'_s}
   >\frac{\alpha_{F_s}}{|\im\F_s|}
 }}
 p_{\zz_s,\zz'_s}
 &=
 \sum_{\substack{
   \zz_s\in\Z^n_s\setminus\{\zz'_s\}:
   \\
   p_{\zz_s,\zz'_s}
   >\frac{\alpha_{F_s}}{|\im\F_s|}
 }}
 p_{\zz_s,\zz'_s}
 +
 p_{\zz'_s,\zz'_s}
 \notag
 \\
 &\leq
 \beta_{F_s}+1
 \label{eq:proof-beta}
\end{align}
for all $\zz'_s\in\Z_s^n$, which comes from (\ref{eq:hash})
and the fact that $p_{\zz'_s,\zz'_s}=1$,

First, we have
\begin{align}
 \sum_{\substack{
   \zz_{\cS}\in\T:
   \\
   p_{\zz_s,\zz'_s}
   >\frac{\alpha_{F_s}}{|\im\F_s|}
   \ \text{for all}\  s\in\cS'
   \\
   p_{\zz_s,\zz'_s}
   \leq\frac{\alpha_{F_s}}{|\im\F_s|}
   \ \text{for all}\ s\in\cS'^{\complement}
 }}
 Q(\zz_{\cS})
 \prod_{s\in\cS} p_{\zz_s,\zz'_s}
 &=
 \sum_{\substack{
   \zz_{\cS'}\in\T_{\cS'}:
   \\
   p_{\zz_s,\zz'_s}
   >\frac{\alpha_{F_s}}{|\im\F_s|}
   \\
   \text{for all}\  s\in\cS'
 }}
 \lrB{\prod_{s\in\cS'} p_{\zz_s,\zz'_s}}
 \sum_{\substack{
   \zz_{\cS'^{\complement}}\in\T_{\cS'^{\complement}|\cS'}\lrsb{\zz_{\cS'}}:
   \\
   p_{\zz_s,\zz'_s}
   \leq\frac{\alpha_{F_s}}{|\im\F_s|}
   \\
   \text{for all}\  s\in\cS'
 }}
 Q(\zz_{\cS'},\zz_{\cS'^{\complement}})
 \prod_{s\in\cS'^{\complement}}p_{\zz_s,\zz'_s}
 \notag
 \\
 &\leq
 \lrB{\prod_{s\in\cS'^{\complement}}\frac{\alpha_{F_s}}{|\im\F_s|}}
 \sum_{\substack{
   \zz_{\cS'}\in\T_{\cS'}:
   \\
   p_{\zz_s,\zz'_s}
   >\frac{\alpha_{F_s}}{|\im\F_s|}
   \\
   \text{for all}\  s\in\cS'
 }}
 \lrB{
  \prod_{s\in\cS'} p_{\zz_s,\zz'_s}
 }
 \sum_{\zz_{\cS'^{\complement}}\in\T_{\cS'^{\complement}|\cS'}\lrsb{\zz_{\cS'}}}
 Q(\zz_{\cS'},\zz_{\cS'^{\complement}})
 \notag
 \\
 &\leq
 \oQ_{\cS'^{\complement}}
 \lrB{\prod_{s\in\cS'^{\complement}}\frac{\alpha_{F_s}}{|\im\F_s|}}
 \prod_{s\in\cS'}
 \lrB{
  \sum_{\substack{
    \zz_s\in\Z_s^n:
    \\
    p_{\zz_s,\zz'_s}
    >\frac{\alpha_{F_s}}{|\im\F_s|}
  }}
  p_{\zz_s,\zz'_s}
 }
 \notag
 \\
 &\leq
 \oQ_{\cS'^{\complement}}
 \lrB{\prod_{s\in\cS'^{\complement}}\frac{\alpha_{F_s}}{|\im\F_s|}}
 \prod_{s\in\cS'}
 \lrB{\beta_{F_s}+1}
 \notag
 \\
 &=
 \frac{
  \alpha_{F_{\cS'^{\complement}}}\lrB{\beta_{F_{\cS'}}+1}
  \oQ_{\cS'^{\complement}}
 }
 {\prod_{s\in\cS'^{\complement}}\lrbar{\im\F_s}}
 \label{eq:lemma-multi-subset}
\end{align}
for all $(\zz'_{\cS},\cS')$ satisfying
$\zz'_{\cS}\in\T$ and $\emptyset\neq\cS'\subsetneq\cS$,
where the second inequality comes from (\ref{eq:maxQJ})
and the third inequality comes from (\ref{eq:proof-beta}).
It should be noted that (\ref{eq:lemma-multi-subset})
is valid for the cases of
$\cS'^{\complement}=\emptyset$ and $\cS'^{\complement}=\cS$
by letting $\oQ_{\emptyset}\equiv Q(\T)$
because
\begin{align}
 \sum_{\substack{
   \zz_{\cS}\in\T:
   \\
   p_{\zz_s,\zz'_s}
   \leq\frac{\alpha_{F_s}}{|\im\F_s|}
   \ \text{for all}\ s\in\cS
 }}
 Q(\zz_{\cS})
 \prod_{s\in\cS} p_{\zz_s,\zz'_s}
 &\leq
 \frac{\alpha_{F_{\cS}}Q(\T)}
 {\prod_{s\in\cS}\lrbar{\im\F_s}}
 \notag
 \\
 &=
 \frac{\alpha_{F_{\cS}}\lrB{\beta_{F_{\emptyset}}+1}\oQ_{\emptyset}}
 {\prod_{s\in\cS}\lrbar{\im\F_s}}
\end{align}
and
\begin{align}
 \sum_{\substack{
   \zz_{\cS}\in\T:
   \\
   p_{\zz_s,\zz'_s}
   >\frac{\alpha_{F_s}}{|\im\F_s|}
   \ \text{for all}\ s\in\cS
 }}
 Q(\zz_{\cS})
 \prod_{s\in\cS} p_{\zz_s,\zz'_s}
 &\leq
 \lrB{\max_{\zz_{\cS}\in\T}Q(\zz_{\cS})}
 \sum_{\substack{
   \zz_{\cS}\in\T:
   \\
   p_{\zz_s,\zz'_s}
   >\frac{\alpha_{F_s}}{|\im\F_s|}
   \ \text{for all}\ s\in\cS
 }}
 \prod_{s\in\cS} p_{\zz_s,\zz'_s}
 \notag
 \\
 &\leq
 \lrB{\max_{\zz_{\cS}\in\T}Q(\zz_{\cS})}
 \prod_{s\in\cS}
 \lrB{
  \sum_{\substack{
    \zz_s\in\Z_s^n:
    \\
    p_{\zz_s,\zz'_s}
    >\frac{\alpha_{F_s}}{|\im\F_s|}
  }}
  p_{\zz_s,\zz'_s}
 }
 \notag
 \\
 &\leq
 \lrB{\max_{\zz_{\cS}\in\T}Q(\zz_{\cS})}
 \prod_{s\in\cS}
 \lrB{\beta_{F_s}+1}
 \notag
 \\
 &=
 \frac{\alpha_{F_{\emptyset}}\lrB{\beta_{F_{\cS}}+1}\oQ_{\cS}}
 {\prod_{s\in\emptyset}\lrbar{\im\F_s}}.
 \label{eq:proof-BCP-multi1}
\end{align}
Then we have
\begin{align}
 \sum_{\zz_{\cS}\in\T}Q(\zz_{\cS})
 \prod_{s\in\cS}p_{\zz_s,\zz'_s}
 &\leq
 \sum_{\cS'\subset\cS}
 \sum_{\substack{
   \zz_{\cS}\in\T:
   \\
   p_{\zz_s,\zz'_s}
   >\frac{\alpha_{F_s}}{|\im\F_s|}
   \ \text{for all}\ s\in\cS'
   \\
   p_{\zz_s,\zz'_s}
   \leq\frac{\alpha_{F_s}}{|\im\F_s|}
   \ \text{for all}\ s\in\cS'^{\complement}
 }}
 Q(\zz_{\cS})
 \prod_{s\in\cS} p_{\zz_s,\zz'_s}
 \notag
 \\
 &\leq
 \sum_{\cS'\subset\cS}
 \frac{
  \alpha_{F_{\cS'^{\complement}}}\lrB{\beta_{F_{\cS'}}+1}
  \oQ_{\cS'^{\complement}}
 }{\prod_{s\in\cS'^{\complement}}\lrbar{\im\F_s}}
 \notag
 \\
 &
 =
 \frac{\alpha_{F_{\cS}}Q(\T)}
 {\prod_{s\in\cS}\lrbar{\im\F_s}}
 +
 \sum_{\substack{
   \cS'\subset\cS:
   \\
   \cS'\neq\emptyset
 }}
 \frac{
  \alpha_{F_{\cS'^{\complement}}}\lrB{\beta_{F_{\cS'}}+1}
  \oQ_{\cS'}
 }
 {\prod_{s\in\cS'^{\complement}}\lrbar{\im\F_s}}
 \label{eq:proof-BCP-multi2}
\end{align}
for all $\zz'_{\cS}\in\T$,
where the equality comes from the fact that $\oQ_{\emptyset}=Q(\T)$ and
$\beta_{F_{\emptyset}}=0$.

Next, let $C_{\cS}$ be the random variable subject to the uniform
distribution on $\Prod_{s\in\cS}\im\F_s$.
From (\ref{eq:proof-BCP-multi2}), we have
\begin{align}
 E_{F_{\cS}C_{\cS}}\lrB{
  \lrB{
   \sum_{\zz_{\cS}\in\T}Q(\zz_{\cS})\chi(F_{\cS}(\zz_{\cS})=C_{\cS})
  }^2
 }
 &=
 \sum_{\zz'_{\cS}\in\T}Q(\zz'_{\cS})
 \sum_{\zz_{\cS}\in\T}Q(\zz_{\cS})
 E_{F_{\cS}}\lrB{
  \chi(F_{\cS}(\zz_{\cS})=F_{\cS}(\zz'_{\cS}))
  E_{C_{\cS}}\lrB{\chi(F_{\cS}(\zz_{\cS})=C_{\cS})}
 }
 \notag
 \\
 &=
 \frac1{\prod_{s\in\cS}\lrbar{\im\F_s}}
 \sum_{\zz'_{\cS}\in\T}Q(\zz'_{\cS})
 \sum_{\zz_{\cS}\in\T}Q(\zz_{\cS})
 \prod_{s\in\cS}p_{\zz_s,\zz'_s}
 \notag
 \\
 &\leq
 \frac{\alpha_{F_{\cS}}Q(\T)^2}
 {\lrB{\prod_{s\in\cS}\lrbar{\im\F_s}}^2}
 +
 \frac{Q(\T)}
 {\prod_{s\in\cS}\lrbar{\im\F_s}}
 \sum_{\substack{
   \cS'\subset\cS:
   \\
   \cS'\neq\emptyset
 }}
 \frac{
  \alpha_{F_{\cS'^{\complement}}}\lrB{\beta_{F_{\cS'}}+1}
  {\oQ_{\cS'^{\complement}}}
 }
 {\prod_{s\in\cS'^{\complement}}\lrbar{\im\F_s}}.
 \label{eq:lemma-multi}
\end{align}
Then we have
\begin{align}
 &
 E_{F_{\cS}C_{\cS}}\lrB{
  \lrB{
   \frac{Q\lrsb{\T\cap\fC_{F_{\cS}}(C_{\cS})}
    \prod_{s\in\cS}|\im\F_s|}
   {Q(\T)}
   -1}^2
 }
 \notag
 \\*
 &=
 E_{F_{\cS}C_{\cS}}\lrB{
  \lrB{\sum_{\zz_{\cS}\in\T}
   \frac{Q(\zz)\chi(F_{\cS}(\zz_{\cS})=C_{\cS})
    \prod_{s\in\cS}|\im\F_s|
   }{Q(\T)}
  }^2
 }
 -2
 E_{F_{\cS}C_{\cS}}\lrB{
  \sum_{\zz_{\cS}\in\T}
  \frac{Q(\zz)\chi(F_{\cS}(\zz_{\cS})=C_{\cS})
   \prod_{s\in\cS}|\im\F_s|
  }{Q(\T)}
 }
 +1
 \notag
 \\
 &=
 E_{F_{\cS}C_{\cS}}\lrB{
  \lrB{\sum_{\zz_{\cS}\in\T}
   \frac{Q(\zz)\chi(F_{\cS}(\zz_{\cS})=C_{\cS})
    \prod_{s\in\cS}|\im\F_s|
   }{Q(\T)}
  }^2
 }
 -2
 \sum_{\zz_{\cS}\in\T}
 \frac{
  Q(\zz)
  E_{F_{\cS}C_{\cS}}\lrB{\chi(F_{\cS}(\zz_{\cS})=C_{\cS})}
  \prod_{s\in\cS}|\im\F_s|
 }{Q(\T)}
 +1
 \notag
 \\
 &=
 \frac{\displaystyle
  \lrB{\prod_{s\in\cS}|\im\F_s|}^2
 }{Q(\T)^2}
 E_{F_{\cS}C_{\cS}}\lrB{
  \lrB{\sum_{\zz_{\cS}\in\T}
   Q(\zz_{\cS})\chi(F_{\cS}(\zz_{\cS})=C_{\cS})
  }^2
 }
 -1
 \notag
 \\
 &\leq
 \alpha_{F_{\cS}}-1
 +
  \sum_{\substack{
    \cS'\subset\cS
    \\
    \cS'\neq\emptyset
  }}
  \alpha_{F_{\cS'^{\complement}}}\lrB{\beta_{F_{\cS'}}+1}
  \lrB{\prod_{s\in\cS'}\lrbar{\im\F_s}}
  \cdot
 \frac{
  \oQ_{\cS'^{\complement}}
 }
 {Q(\T)},
\end{align}
where the inequality comes from (\ref{eq:lemma-multi}).

Finally, the lemma is confirmed by
\begin{align}
 E_{F_{\cS}}\lrB{
  \sum_{\cc_{\cS}}
  \left|
   \frac{Q\lrsb{\T\cap\fC_{F_{\cS}}(\cc_{\cS})}}{Q(\T)}
   -\frac 1{\prod_{s\in\cS}|\im\F_s|}
  \right|
 }
 &=
 E_{F_{\cS}C_{\cS}}\lrB{
  \left|
   \frac{Q\lrsb{\T\cap\fC_{F_{\cS}}(C_{\cS})}
    \prod_{s\in\cS}|\im\F_s|
   }{Q(\T)}
   -1
  \right|
 }
 \notag
 \\
 &=
 E_{F_{\cS}C_{\cS}}\lrB{
  \sqrt{
   \lrB{
    \frac{
     Q\lrsb{\T\cap\fC_{F_{\cS}}(C_{\cS})}
     \prod_{s\in\cS}|\im\F_s|
    }{Q(\T)}
    -1}^2
  }
 }
 \notag
 \\
 &\leq
 \sqrt{
  E_{F_{\cS}C_{\cS}}\lrB{
   \lrB{\frac{
     Q\lrsb{\T\cap\fC_{F_{\cS}}(C_{\cS})}
     \prod_{s\in\cS}|\im\F_s|
    }{Q(\T)}
    -1}^2
  }
 }
 \notag
 \\
 &\leq
 \sqrt{
  \alpha_{F_{\cS}}-1
  +
  \!\!\!\!\!\!
   \sum_{\substack{
     \cS'\subset\cS:
     \cS'\neq\emptyset
   }}
  \!\!\!\!\!\!
   \alpha_{F_{\cS'^{\complement}}}
   \lrB{\beta_{F_{\cS}}+1}
   \lrB{\prod_{s\in\cS'}|\im\F_s|}
   \cdot
  \frac{
   \oQ_{\cS'^{\complement}}
  }
  {Q(\T)}
 },
 \label{eq:proof-BCP-multi}
\end{align}
where the first inequality comes from the Jensen inequality.
\end{IEEEproof}

The following lemma is a multiple extension of
the {\it collision-resistant property}.
This lemma implies that
there is an assignment such that every bin contains at most one item.
\begin{lem}[{\cite[Lemma 7 in the extended version]{HASH-BC}}]
\label{lem:mCRP}
For each $s\in\cS$, let $\F_s$ be a set of functions on $\Z_s^n$
and $p_{F_s}$ be the probability distribution on $\F_s$,
where $(\F_s,p_{F_s})$ satisfies (\ref{eq:hash}).
We assume that random variables $\{F_s\}_{s\in\cS}$ are mutually independent.
Then
\begin{align*}
 p_{F_{\cS}}\lrsb{\lrb{
   f_{\cS}:
   \lrB{\T\setminus\{\zz_{\cS}\}}\cap\fC_{f_{\cS}}(f_{\cS}(\zz_{\cS}))
   \neq
   \emptyset
 }}
 \leq
 \sum_{\substack{
   \cS'\subset\cS:
   \\
   \cS'\neq\emptyset
 }}
 \frac{
  \alpha_{F_{\cS'}}\lrB{\beta_{F_{\cS'^{\complement}}}+1}
  \oO_{\cS'}
 }
 {
  \prod_{s\in\cS'}\lrbar{\im\F_s}
 }
 +\beta_{F_{\cS}}
\end{align*}
for all $\T\subset\Z_{\cS}^n$ and $\zz_{\cS}\in\Z_{\cS}^n$,
where
\begin{equation*}
 \oO_{\cS'}
 \equiv
 \begin{cases}
  |\T|
  &\text{if}\ \cS'=\cS,
  \\
  \displaystyle\max_{\zz_{\cS'^{\complement}}\in\T_{\cS'^{\complement}}}
  \lrbar{\T_{\cS'|\cS'^{\complement}}\lrsb{\zz_{\cS'^{\complement}}}},
  &\text{if}\ \emptyset\neq\cS'\subsetneq\cS
 \end{cases}
\end{equation*}
\end{lem}
\begin{IEEEproof}
Let
$p_{\zz_s,\zz'_s}
\equiv
p_{F_s}\lrsb{\lrb{
  f_s:
  f_s(\zz_s)=f_s(\zz'_s)
}}$.
By interchanging $\cS'$ and $\cS'^{\complement}$,
and letting $\oO_{\emptyset}=1$ and $O(\zz_{\cS})\equiv 1$
for each $\zz_{\cS}\in\Z_{\cS}^n$,
we have the fact that
\begin{align}
 \sum_{\substack{
   \zz_{\cS'}\in\T:
   \\
   p_{\zz_s,\zz'_s}
   \leq\frac{\alpha_{F_s}}{|\im\F_s|}
   \ \text{for all}\ s\in\cS'
   \\
   p_{\zz_s,\zz'_s}
   >\frac{\alpha_{F_s}}{|\im\F_s|}
   \ \text{for all}\ s\in\cS'^{\complement}
 }}
 \prod_{s\in\cS'} p_{\zz_s,\zz'_s}
 &\leq
 \frac{\alpha_{F_{\cS'}}\lrB{\beta_{F_{\cS'^{\complement}}}+1}\oO_{\cS'}}
 {\prod_{s\in\cS'}|\im\F_s|}
 \label{eq:lemma-mcrp}
\end{align}
for all $\zz'_{\cS'}\in\Z_{\cS'}^n$ and $\cS'\subset\cS$
from (\ref{eq:lemma-multi-subset}).
Then we have
\begin{align}
 p_{F_{\cS}}\lrsb{\lrb{
   f_{\cS}:
   \lrB{\T\setminus\{\zz_{\cS}\}}\cap\fC_{F_{\cS}}(F_{\cS}\zz_{\cS})
   \neq
   \emptyset
 }}
 &\leq
 \sum_{\zz_{\cS}\in\T\setminus\{\zz'_{\cS}\}}
 p_{F_{\cS}}\lrsb{\lrb{
   f_{\cS}:
   f_{\cS}(\zz_{\cS})=f_{\cS}(\zz'_{\cS})
 }}
 \notag
 \\
 &=
 \sum_{\zz_{\cS}\in\T\setminus\{\zz'_{\cS}\}}
 p_{F_{\cS}}\lrsb{\lrb{
   f_{\cS}:
   f_s(\zz_s)=f_s(\zz'_s)
   \ \text{for all}\ s\in\cS
 }}
 \notag
 \\*
 &=
 \sum_{\zz_{\cS}\in\T}
 \prod_{s\in\cS} p_{\zz_s,\zz'_s}
 -
 \prod_{s\in\cS} p_{\zz'_s,\zz'_s}
 \notag
 \\
 &=
 \sum_{\cS'\subset\cS}
 \sum_{\substack{
   \zz_{\cS}\in\T:
   \\
   p_{\zz_s,\zz'_s}
   \leq\frac{\alpha_{F_s}}{|\im\F_s|}
   \ \text{for all}\ s\in\cS'
   \\
   p_{\zz_s,\zz'_s}
   >\frac{\alpha_{F_s}}{|\im\F_s|}
   \ \text{for all}\ s\in\cS'^{\complement}
 }}
 \prod_{s\in\cS} p_{\zz_s,\zz'_s}
 -1
 \notag
 \\
 &\leq
 \sum_{\cS'\subset\cS}
 \frac{\alpha_{F_{\cS'}}\lrB{\beta_{F_{\cS'^{\complement}}}+1}\oO_{\cS'}}
 {\prod_{s\in\cS}|\im\F_s|}
 -1
 \notag
 \\
 &=
 \sum_{\substack{
   \cS'\subset\cS:
   \\
   \cS'\neq\emptyset
 }}
 \frac{
  \alpha_{F_{\cS'}}\lrB{\beta_{F_{\cS'^{\complement}}}+1}\oO_{\cS'}
 }
 {\prod_{s\in\cS'}|\im\F_s|}
 +
 \beta_{F_{\cS}}
\end{align}
for all $\T\subset\Z_{\cS'}^n$ and $\zz'_{\cS'}\in\Z_{\cS'}^n$,
where the third equality comes from the fact that $p_{\zz'_s,\zz'_s}=1$,
the second inequality comes from (\ref{eq:lemma-mcrp}),
and the last equality comes from the fact that
$\alpha_{F_{\emptyset}}=1$,
$\beta_{F_{\emptyset^{\complement}}}=\beta_{F_{\cS}}$,
$\prod_{s\in\emptyset}|\im\F_s|=1$, and $\oO_{\emptyset}=1$.
\end{IEEEproof}

\subsection{Proof of Lemma~\ref{lem:channel-bcp}}
\label{sec:channel-bcp}

Let us assume that
ensembles $(\F_s,p_{F_s})$ and $(\G_s,p_{G_s})$ have the hash property 
((\ref{eq:hash}) in Appendix \ref{sec:hash})
for every $s\in\cS$,
where their dependence on $n$ is omitted.
In the following, we omit the dependence of $Z$ on $n$,
when it appears in the subscript of $\mu$.
Moreover, we omit the dependence of $\alpha$ and $\beta$ on $n$,

From Lemma \ref{lem:hash-FG} in Appendix \ref{sec:hash},
we have the fact that
the joint ensemble $(\F_s\times\G_s,p_{(F,G)_s})$
also satisfies the hash property.
In the proof of Theorem \ref{thm:channel},
we apply the following lemma
to the joint ensemble $(\F_s\times\G_s,p_{(F,G)_s})$.
\begin{lem}[{\cite[Eq. (50)]{ICC}}]
\label{lem:channel-bcp}
For given disjoint sets $\cS$ and $\ocS$,
let $\{Z^n_{\cS\cup\ocS}\}_{n=1}^{\infty}$ be general correlated
sources, where $Z^n_{\cS\cup\ocS}\equiv\{Z^n_s\}_{s\in\cS\cup\ocS}$.
Let $\uT$ be defined as
\begin{equation*}
 \uT
 \equiv
 \lrb{
  (\zz_{\ocS},\zz_{\cS}):
  \begin{aligned}
   &\frac 1n
   \log_2\frac1{
    \mu_{Z_{\cS'}|Z_{\ocS}}(\zz_{\cS'}|\zz_{\ocS})
   }
   \geq
   \uH(\ZZ_{\cS'}|\ZZ_{\ocS})-\e
   \\
   &\text{for all}\ \cS'\subset\cS\cup\ocS
  \end{aligned}
 }.
\end{equation*} 
Then we have
\begin{align*}
 &
 E_{F_{\cS}}\lrB{
  \sum_{
   \zz_{\ocS}\in\Z^n_{\ocS},
   \cc_{\cS}\in\im\F_{\cS}
  }
  \mu_{Z_{\ocS}}(\zz_{\ocS})
  \lrbar{
   \mu_{Z_{\cS}|Z_{\ocS}}(
    \fC_{F_{\cS}}(\cc_{\cS})
    |\zz_{\ocS}
   )
   -
   \frac{
    1
   }{
    \prod_{s\in\cS}
    |\im\F_s|
   }
  }
 }
 \\*
 &\leq
 \sqrt{
  \alpha_{F_{\cS}}-1
  +\sum_{
    \cS'\subset\cS:
    \cS'\neq\emptyset
  }
  \alpha_{F_{\cS\setminus\cS'}}
  [\beta_{F_{\cS'}}+1]
  \lrB{
   \prod_{s\in\cS'}|\im\F_s|
  }
  2^{-n[\uH(\ZZ_{\cS'}|\ZZ_{\ocS})-\e]}
 }
 +
 2\mu_{Z_{\ocS\cup\cS}}(\uT^{\complement}).
\end{align*}
\end{lem}

\begin{IEEEproof}
Let $\uT(\zz_{\ocS})$ be defined as
\begin{equation*}
 \uT(\zz_{\ocS})
 \equiv
 \lrb{
  \zz_{\cS}:
  (\zz_{\ocS},\zz_{\cS})
  \in\uT
 }.
\end{equation*}
Then we have
\begin{align}
 &
 E_{F_{\cS}}\lrB{
  \sum_{
   \zz_{\ocS}\in\Z^n_0,
   \cc_{\cS}\in\im\F_{\cS}
  }
  \mu_{Z_{\ocS}}(\zz_{\ocS})
  \lrbar{
   \mu_{Z_{\cS}|Z_{\ocS}}(
    \fC_{F_{\cS}}(\cc_{\cS})
    |\zz_{\ocS}
   )
   -
   \frac{
    1
   }{
    \prod_{s\in\cS}
    |\im\F_s|
   }
  }
 }
 \notag
 \\*
 &\leq
 E_{F_{\cS}}
 \lrB{
  \sum_{\substack{
    \zz_{\ocS}\in\Z^n_0,
    \cc_{\cS}\in\im\F_{\cS}
  }}
  \mu_{Z_{\ocS}}(\zz_{\ocS})
  \lrbar{
   \mu_{Z_{\cS}|Z_{\ocS}}(
    \uT(\zz_{\ocS})
    \cap\fC_{F_{\cS}}(\cc_{\cS})
    |\zz_{\ocS}
   )
   -
   \frac{
    \mu_{Z_{\cS}|Z_{\ocS}}(
     \uT(\zz_{\ocS})
     |\zz_{\ocS}
    )
   }{
    \prod_{s\in\cS}
    |\im\F_s|
   }
  }
 }
 \notag
 \\*
 &\quad
 +
 E_{F_{\cS}}
 \lrB{
  \sum_{\substack{
    \zz_{\ocS}\in\Z^n_0,
    \cc_{\cS}\in\im\F_{\cS}
  }}
  \mu_{Z_{\cS}|Z_{\ocS}}(
   \uT(\zz_{\ocS})^{\complement}
   \cap\fC_{F_{\cS}}(\cc_{\cS})
   |\zz_{\ocS}
  )
  \mu_{Z_{\ocS}}(\zz_{\ocS})
 }
 \notag
 \\*
 &\quad
 +
 E_{F_{\cS}}
 \lrB{
  \sum_{\substack{
    \zz_{\ocS}\in\Z^n_{\ocS},
    \cc_{\cS}\in\im\F_{\cS}
  }}
  \frac{
   \mu_{Z_{\cS}|Z_{\ocS}}(
    \uT_{Z_{\cS}|Z_{\ocS}}(\zz_{\ocS})^{\complement}
    |\zz_{\ocS}
   )
   \mu_{Z_{\ocS}}(\zz_{\ocS})
  }{
    \prod_{s\in\cS}
    |\im\F_s|
  }
 }
 \notag
 \\
 &=
 \sum_{
  \zz_{\ocS}\in\Z^n_{\ocS}
 }
 \mu_{Z_{\cS}|Z_{\ocS}}(
  \uT(\zz_{\ocS})
  |\zz_{\ocS}
 )
 \mu_{Z_{\ocS}}(\zz_{\ocS})
 E_{F_{\cS}}\lrB{
  \sum_{\substack{
    \cc_{\cS}\in\im\F_{\cS}
  }}
  \lrbar{
   \frac{
    \mu_{Z_{\cS}|Z_{\ocS}}(
     \uT(\zz_{\ocS})
     \cap\fC_{F_{\cS}}(\cc_{\cS})
     |\zz_{\ocS}
    )
   }{
    \mu_{Z_{\cS}|Z_{\ocS}}(
     \uT(\zz_{\ocS})
     |\zz_{\ocS}
    )
   }
   -
   \frac1
   {\prod_{s\in\cS}|\im\F_s|}
  }
 }
 \notag
 \\*
 &\quad
 +
 2
 \sum_{
  \zz_{\ocS}\in\Z^n_{\ocS}
 }
 \mu_{Z_{\cS}|Z_{\ocS}}(
  \uT(\zz_{\ocS})^{\complement}
  |\zz_{\ocS}
 )
 \mu_{Z_{\ocS}}(\zz_{\ocS})
 \notag
 \\
 &\leq
 \sum_{
  \zz_{\ocS}\in\Z^n_{\ocS}
 }
 \mu_{Z_{\cS}|Z_{\ocS}}(
  \uT(\zz_{\ocS})
  |\zz_{\ocS}
 )
 \mu_{Z_{\ocS}}(\zz_{\ocS})
 \sqrt{
  \alpha_{F_{\cS}}-1
  +
  \sum_{
    \cS'\subset\cS:
    \cS'\neq\emptyset
  }
  \alpha_{F_{\cS\setminus\cS'}}
  [\beta_{F_{\cS'}}+1]
  \lrB{
   \prod_{s\in\cS'}|\im\F_s|
  }
  \frac{
   2^{-n[\uH(\ZZ_{\cS'}|\ZZ_{\ocS})-\e]}
  }{
   \mu_{Z_{\cS}|Z_{\ocS}}(
    \uT(\zz_{\ocS})
    |\zz_{\ocS}
   )
  }
 }
 \notag
 \\*
 &\quad
 +
 2\mu_{Z_{\ocS\cup\cS}}(\uT^{\complement}),
 \notag
 \\
 &\leq
 \sum_{
  \zz_{\ocS}\in\Z^n_{\ocS}
 }
 \mu_{Z_{\ocS}}(\zz_{\ocS})
 \sqrt{
  \alpha_{F_{\cS}}-1
  +
  \sum_{
    \cS'\subset\cS:
    \cS'\neq\emptyset
  }
  \alpha_{F_{\cS\setminus\cS'}}
  [\beta_{F_{\cS'}}+1]
  \lrB{
   \prod_{s\in\cS'}|\im\F_s|
  }
  2^{-n[\uH(\ZZ_{\cS'}|\ZZ_{\ocS})-\e]}
 }
 +
 2\mu_{Z_{\ocS\cup\cS}}(\uT^{\complement})
 \notag
 \\
 &=
 \sqrt{
  \alpha_{F_{\cS}}-1
  +\sum_{
    \cS'\subset\cS:
    \cS'\neq\emptyset
  }
  \alpha_{F_{\cS\setminus\cS'}}
  [\beta_{F_{\cS'}}+1]
  \lrB{
   \prod_{s\in\cS'}|\im\F_s|
  }
  2^{-n[\uH(\ZZ_{\cS'}|\ZZ_{\ocS})-\e]}
 }
 +
 2\mu_{Z_{\ocS\cup\cS}}(\uT^{\complement}),
 \label{eq:proof-crng-bcp-5}
\end{align}
where the second inequality comes from
Lemma~\ref{lem:mBCP} in Appendix \ref{sec:hash}
by letting
\begin{align*}
 \T
 &\equiv
 \uT(\zz_{\ocS})
 \\
 Q
 &\equiv
 \mu_{Z_{\cS}|Z_{\ocS}}(\cdot|\zz_{\ocS})
\end{align*}
and using the relations
\begin{align}
 \T_{\cS'}
 &\subset
 \lrb{
  \zz_{\cS'}:
  \frac 1n
  \log_2\frac1{\mu_{Z_{\cS'}|Z_{\ocS}}(\zz_{\cS'}|\zz_{\ocS})}
  \geq
  \uH(\ZZ_{\cS'}|\ZZ_{\ocS})-\e
 }
 \notag
 \\
 \oQ_{\cS'^{\complement}}
 &=
 \max_{\zz_{\cS'}\in\T_{\cS'}}
 \sum_{\zz_{\cS'^{\complement}}
  \in\T_{\cS'^{\complement}|\cS'}(\zz_{\cS'})}
 \mu_{Z_{\cS}|Z_{\ocS}}
 (\zz_{\cS}|\zz_{\ocS})
 \notag
 \\
 &\leq
 \max_{\zz_{\cS'}\in\T_{\cS'}}
 \mu_{Z_{\cS'}|Z_{\ocS}}(\zz_{\cS'}|\zz_{\ocS})
 \notag
 \\
 &\leq
 2^{-n[\uH(\ZZ_{\cS'}|\ZZ_{\ocS})-\e]}.
\end{align}
\end{IEEEproof}

\subsection{Proof of Lemma~\ref{lem:channel-crp}}
\label{sec:channel-crp}

Assume that $(\F_s,p_{F_s})$ has the hash property
((\ref{eq:hash}) in Appendix \ref{sec:hash})
for every $s\in\D_j$,
where their dependence on $n$ is omitted.
In the following, we also omit the dependence of $C$, $Y$, $Z$, and $\hZ$ on $n$,
when it appears in the subscript of $\mu$.
We omit the dependence of $\alpha$ and $\beta$ on $n$.
In the following, we fix $j\in\J$ and omit subscript $j$.

Let us assume that $(\ZZ_{\cS},\CC_{\cS},\YY,\hZZ_{\D})$ satisfies
the Markov relation
\begin{equation}
 Z^n_{\cS}\markov(C^{(n)}_{\cS},Y^n)\markov\hZ^n_{\D}
 \label{eq:markov-decoding}
\end{equation}
and
\begin{equation}
 f_s(Z^n_s)=C^{(n)}_s
 \label{eq:source-encoding}
\end{equation}
for all $s\in\cS$ and $n\in\NN$.
For given $\e>0$, let $\oT$ be defined as
\begin{equation}
 \oT
 \equiv
 \lrb{
  (\zz_{\D},\yy):
  \begin{aligned}
   &
   \frac1n\log_2
   \frac1{\mu_{Z_{\D'}|YZ_{\D\setminus\D'}}
    (\zz_{\D'}|\yy,\zz_{\D\setminus\D'})}
   \leq
   \oH(\ZZ_{\D'}|\YY,\ZZ_{\D\setminus\D'})+\e
   \\
   &\text{for all}\ \D'\ \text{satisfying}\ \emptyset\neq\D'\subset\D
  \end{aligned}
 }.
 \label{eq:oT}
\end{equation}

First, we show the following lemma.
\begin{lem}[{\cite[Eq. (58)]{CRNG-MULTI}}]
\label{lem:tsdecoding}
Let us assume that $\chzz_{\D}(\cc_{\D}|\yy)$ outputs one of the elements
in $\oT\cap\fC_{f_{\D}}(\cc_{\D})$
and declares an error when $\oT\cap\fC_{f_{\D}}(\cc_{\D})=\emptyset$.
Then we have
\begin{align*}
 &
 E_{F_{\D}}\lrB{
  \mu_{Z_{\D}Y}\lrsb{
   \lrb{
    (\zz_{\D},\yy): \chzz_{\D}(F_{\D}(\zz_{\D})|\yy)\neq\zz_{\D}
   }
  }
 }
 \notag
 \\*
 &\leq
 \sum_{
  \D'\subset\D:
  \D'\neq\emptyset
 }
 \alpha_{F_{\D'}}\lrB{\beta_{F_{\D\setminus\D'}}+1}
 \frac{
  2^{
   n\lrB{\oH(\ZZ_{\D'}|\YY,\ZZ_{\D\setminus\D'})+\e}
  }
 }{
  \prod_{s\in\D'}|\im\F_s|
 }
 +\beta_{F_{\D}}
 +\mu_{Z_{\D}Y}(\oT^{\complement}).
\end{align*}
\end{lem}
\begin{IEEEproof}
 Let $\oT(\yy)\equiv\{\zz_{\D}: (\zz_{\D},\yy)\in\oT\}$
 and assume that $(\zz_{\D},\yy)\in\oT$
 and $\chzz_{\D}(f_{\D}(\zz_{\D})|\yy)\neq\zz_{\D}$.
 Since $\zz_{\D}\in\fC_{f_{\D}}(f_{\D}(\zz_{\D}))$, we have
 $\lrB{\oT(\yy)\setminus\{\zz_{\D}\}}
 \cap\fC_{f_{\D}}(f_{\D}(\zz_{\D}))\neq\emptyset$.
 We have
 \begin{align}
  E_{F_{\D}}\lrB{
   \chi(\chzz_{\D}(F_{\D}(\zz_{\D})|\yy)\neq\zz_{\D})
  }
  &\leq
  p_{F_{\D}}\lrsb{\lrb{
    f_{\D}:
    \lrB{\oT_{\Z_{\D}}(\yy)\setminus\{\zz_{\D}\}}
    \cap\C_{f_{\D}}(f_{\D}(\zz_{\D}))\neq\emptyset
  }}
  \notag
  \\
  &\leq
  \sum_{
   \D'\subset\D:
   \D'\neq\emptyset
  }
  \frac{
   \alpha_{F_{\D'}}\lrB{\beta_{F_{\D\setminus\D'}}+1}
   \oO_{\D'}
  }{
   \prod_{s\in\D'}\lrbar{\im\F_s}
  }
  +\beta_{F_{\D}}
  \notag
  \\
  &\leq
  \sum_{
   \D'\subset\D:
   \D'\neq\emptyset
  }
  \alpha_{F_{\D'}}\lrB{\beta_{F_{\D\setminus\D'}}+1}
  \frac{
   2^{
    n\lrB{\oH(\ZZ_{\D'}|\YY,\ZZ_{\D\setminus\D'})+\e}
   }
  }{
   \prod_{s\in\D'}\lrbar{\im\F_s}
  }
  +\beta_{F_{\D}},
 \end{align}
 where
 the second inequality comes from Lemma~\ref{lem:mCRP} 
 in Appendix \ref{sec:hash}
 by letting $\T\equiv\oT$
 and the third inequality comes from the fact that
 \begin{equation*}
  \oO_{\D'}
  \leq
  2^{n\lrB{\oH(\ZZ_{\D'}|\YY,\ZZ_{\D\setminus\D'})+\e}}.
 \end{equation*}
 We have
 \begin{align}
  &
  E_{F_{\D}}\lrB{
   \mu_{Z_{\D}Y}\lrsb{
    \lrb{
     (\zz_{\D},\yy): \chzz_{\D}(F_{\D}(\zz_{\D})|\yy)\neq\zz_{\D}
    }
   }
  }
  \notag
  \\*
  &=
  E_{F_{\D}}\lrB{
   \sum_{
    \zz_{\D}\in\Z^n_{\D},
    \yy\in\Y^n
   }
   \mu_{Z_{\D}Y}(\zz_{\D},\yy)
   \chi(\chzz_{\D}(F_{\D}(\zz_{\D})|\yy)\neq \zz_{\D})
  }
  \notag
  \\*
  &=
  \sum_{(\zz_{\D},\yy)\in\oT_{Z_{\D}}}
  \mu_{Z_{\D}Y}(\zz_{\D},\yy)
  E_{F_{\D}}\lrB{
   \chi(\chzz_{\D}(F_{\D}(\zz_{\D})|\yy)\neq\zz_{\D})
  }
  +
  \sum_{(\zz_{\D},\yy)\in\oT_{Z_{\D}}^{\complement}}
  \mu_{Z_{\D}Y}(\zz_{\D},\yy)
  E_{F_{\D}}\lrB{
   \chi(\chzz_{\D}(F_{\D}(\zz_{\D})|\yy)\neq \zz_{\D})
  }
  \notag
  \\
  &\leq
  \sum_{
   \D'\subset\D:
   \D'\neq\emptyset
  }
  \alpha_{F_{\D'}}\lrB{\beta_{F_{\D\setminus\D'}}+1}
  \frac{
   2^{
    n\lrB{\oH(\ZZ_{\D'}|\YY,\ZZ_{\D\setminus\D'})+\e}
   }
  }{
   \prod_{s\in\D'}\lrbar{\im\F_s}
  }
  +\beta_{F_{\D}}
  +\mu_{Z_{\D}Y}(\oT^{\complement}).
 \end{align}
\end{IEEEproof}

Next, we show the following lemma.
\begin{lem}[{\cite[Lemma 2 in the extended version]{ICC}}]
\label{lem:channel-crp}
The expectation of the decoding error probability is evaluated as follows
\begin{align}
 &
 E_{F_{\D}}\lrB{
  \mu_{Z_{\D}\hZ_{\D}}\lrsb{
   \lrb{
    (\zz_{\D},\hzz_{\D}):
    \hzz_{\D}\neq\zz_{\D}
   }
  }
 }
 \notag
 \\*
 &\leq
 2
 \sum_{
  \D'\subset\D:
  \D'\neq\emptyset
 }
 \alpha_{F_{\D'}}\lrB{\beta_{F_{\D\setminus\D'}}+1}
 2^{
  -n\lrB{
   \sum_{s\in\D'}r_s-\oH(\ZZ_{\D'}|\YY,\ZZ_{\D\setminus\D'})
   -\e
  }
 }
 +2\beta_{F_{\D}}
 +2\mu_{Z_{\D}Y}(\oT^{\complement}),
 \label{eq:source-error}
\end{align}
where $\C_s\equiv\im\F_s$ and the decoding error probability
$\mu_{Z_{\D}\hZ_{\D}}\lrsb{
 \lrb{
  (\zz_{\D},\hzz_{\D}):
  \hzz_{\D}\neq\zz_{\D}
 }
}$
depends on $f_{\D}$ through the relation (\ref{eq:source-encoding}).
\end{lem}
\begin{IEEEproof}
For given $f_{\D}$,
the joint distribution of $(Z_{\D}^n,C^{(n)}_{\D},Y^n)$ is given as
\begin{equation*}
 \mu_{Z_{\D}C_{\D}Y}(\zz_{\D},\cc_{\D},\yy)
 =
 \mu_{Z_{\D}Y}(\zz_{\D},\yy)
 \chi(f_{\D}(\zz_{\D})=\cc_{\D}).
\end{equation*}
Then we have
\begin{align}
 \mu_{Z_{\D}|C_{\D}Y}(\zz_{\D}|\cc_{\D},\yy)
 &\equiv
 \frac{
  \mu_{Z_{\D}C_{\D}Y}(\zz_{\D},\cc_{\D},\yy)
 }{
  \sum_{\zz_{\D}\in\Z^n_{\D}}
  \mu_{Z_{\D}C_{\D}Y}(\zz_{\D},\cc_{\D},\yy)
 }
 \notag
 \\
 &=
 \frac{
  \mu_{Z_{\D}Y}(\zz_{\D},\yy)\chi(f_{\D}(\zz_{\D})=\cc_{\D})
 }{
  \sum_{\zz_{\D}\in\Z^n_{\D}}
  \mu_{Z_{\D}Y}(\zz_{\D},\yy)
  \chi(f_{\D}(\zz_{\D})=\cc_{\D})
 }
 \notag
 \\
 &=
 \frac{
  \mu_{Z_{\D}|Y}(\zz_{\D}|\yy)\chi(f_{\D}(\zz_{\D})=\cc_{\D})
 }{
  \sum_{\zz_{\D}\in\Z^n_{\D}}
  \mu_{Z_{\D}|Y}(\zz_{\D}|\yy)
  \chi(f_{\D}(\zz_{\D})=\cc_{\D})
 }
 \notag
 \\
 &=
 \mu_{\hZ_{\D}|C_{\D}Y}(\zz_{\D}|\cc_{\D},\yy),
\end{align}
that is, the constrained-random-number generator defined by
(\ref{eq:decoder}) is a stochastic decision with $\mu_{Z_{\D}|C_{\D}Y}$.
By letting
\begin{equation*}
 \mu_{\chZ_{\D}|C_{\D}Y}(\chzz_{\D}|\cc_{\D},\yy)
 \equiv
 \chi(\chzz_{\D}(\cc_{\D}|\yy)=\chzz_{\D}),
\end{equation*}
we have the fact that
\begin{align}
 \mu_{Z_{\D}\hZ_{\D}}\lrsb{
  \lrb{
   (\zz_{\D},\hzz_{\D}):
   \hzz_{\D}\neq\zz_{\D}
  }
 }
 &=
 \sum_{\substack{
   \zz_{\D}\in\Z^n_{\D},
   \cc_{\D}\in\C_{\D},
   \\
   \yy\in\Y^n,
   \hzz_{\D}\in\Z^n_{\D}:
   \\
   \hzz_{\D}\neq\zz_{D}
 }}
 \mu_{Z_{\D}C_{\D}Y\hZ_{\D}}(\zz_{\D},\cc_{\D},\yy,\hzz_{\D})
 \notag
 \\
 &=
 \sum_{\substack{
   \zz_{\D}\in\Z^n_{\D},
   \cc_{\D}\in\C_{\D},
   \\
   \yy\in\Y^n,
   \hzz_{\D}\in\Z^n_{\D}:
   \\
   \hzz_{\D}\neq\zz_{D}
 }}
 \mu_{Z_{\D}C_{\D}Y}(\zz_{\D},\cc_{\D},\yy)
 \mu_{Z_{\D}|C_{\D}Y}(\hzz_{\D}|\cc_{\D},\yy)
 \notag
 \\
 &\leq
 2
 \sum_{\substack{
   \zz_{\D}\in\Z^n_{\D},
   \cc_{\D}\in\C_{\D},
   \\
   \yy\in\Y^n,
   \chzz_{\D}\in\Z^n_{\D}:
   \\
   \chzz_{\D}\neq\zz_{D}
 }}
 \mu_{Z_{\D}C_{\D}Y}(\zz_{\D},\cc_{\D},\yy)
 \mu_{\chZ_{\D}|C_{\D}Y}(\chzz_{\D}|\cc_{\D},\yy)
 \notag
 \\
 &=
 2
 \sum_{\substack{
   \zz_{\D}\in\Z^n_{\D},
   \cc_{\D}\in\C_{\D},
   \\
   \yy\in\Y^n,
   \chzz_{\D}\in\Z^n_{\D}:
   \\
   \chzz_{\D}\neq\zz_{D}
 }}
 \mu_{Z_{\D}Y}(\zz_{\D},\yy)
 \chi(f_{\D}(\zz_{\D})=\cc_{\D})
 \chi(\chzz_{\D}(\cc_{\D}|\yy)=\chzz_{\D})
 \notag
 \\
 &=
 2
 \sum_{\substack{
   \zz_{\D}\in\Z^n_{\D},
   \yy\in\Y^n:
   \\
   \chzz_{\D}(f_{\D}(\zz_{\D})|\yy)\neq\zz_{\D}
 }}
 \mu_{Z_{\D}Y}(\zz_{\D},\yy)
 \notag
 \\
 &=
 2
 \mu_{Z_{\D}Y}
 \lrsb{
  \lrb{
   (\zz_{\D},\yy):
   \chzz_{\D}(f_{\D}(\zz_{\D})|\yy)\neq\zz_{\D}
  }
 },
\end{align}
where the second equality comes from (\ref{eq:markov-decoding}),
the first inequality comes from 
Lemma~\ref{lem:sdecoding} in Appendix \ref{sec:proof-lemma},
and the third equality comes from (\ref{eq:source-encoding}).
Then we have the fact that
\begin{align}
 &
 E_{F_{\D}}\lrB{
  \mu_{Z_{\D}\hZ_{\D}}\lrsb{
   \lrb{
    (\zz_{\D},\hzz_{\D}):
    \hzz_{\D}\neq\zz_{\D}
   }
  }
 }
 \notag
 \\*
 &\leq
 2
 E_{F_{\D}}\lrB{
  \mu_{Z_{\D}Y}
  \lrsb{
   \lrb{
    (\zz_{\D},\yy):
    \chzz_{\D}(F_{\D}(\zz_{\D})|\yy)\neq\zz_{\D}
   }
  }
 }
 \notag
 \\
 &\leq
 2 \sum_{
  \D'\subset\D:
  \D'\neq\emptyset
 }
 \alpha_{F_{\D'}}\lrB{\beta_{F_{\D\setminus\D'}}+1}
 2^{
  -n\lrB{
   \sum_{s\in\D'}r_s-\oH(\ZZ_{\D'}|\YY,\ZZ_{\D\setminus\D'})-\e
  }
 }
 +2 \beta_{F_{\D}}
 +2 \mu_{Z_{\D}Y}(\oT^{\complement}),
\end{align}
where we use the relation $r_s=\log_2(|\C_s|)/n=\log_2(|\im\F_s|)/n$
in the last inequality.
\end{IEEEproof}

\subsection{Proof of Lemmas}
\label{sec:proof-lemma}

\begin{lem}
\label{lem:diff-prod}
For any sequence $\{\theta_k\}_{k=1}^K$ of positive numbers,
we have
\begin{equation}
 \lrbar{
  \prod_{k=1}^K \theta_k - 1
 }
 \leq
 \sum_{k=1}^K
 \lrbar{\theta_k - 1}
 \prod_{k'=k+1}^K\theta_{k'},
 \label{eq:diff-prod}
\end{equation}
where $\prod_{k'=K+1}^{K}\theta_{k'}\equiv1$.
\end{lem}
\begin{IEEEproof}
 When $K=1$, (\ref{eq:diff-prod}) is trivial.
 Assume that (\ref{eq:diff-prod}) is satisfied,
 then we have
\begin{align}
 \lrbar{\prod_{k=1}^{K+1} \theta_k - 1}
 &\leq
 \lrbar{
  \prod_{k=1}^{K+1} \theta_k
  -
  \theta_{K+1}
 }
 +
 \lrbar{
  \theta_{K+1}
  - 1
 }
 \notag
 \\
 &=
 \lrbar{
  \prod_{k=1}^K \theta_k
  -
  1
 }
 \theta_{K+1}
 +
 \lrbar{
  \theta_{K+1}
  - 1
 }
 \notag
 \\
 &\leq
 \lrB{
  \sum_{k=1}^K
  \lrbar{\theta_k - 1}
  \prod_{k'=k+1}^K\theta_{k'}
 }
 \theta_{K+1}
 +
 \lrbar{
  \theta_{K+1}
  - 1
 }
 \notag
 \\
 &=
 \sum_{k=1}^{K+1}
 \lrbar{\theta_k - 1}
 \prod_{k'=k+1}^{K+1}\theta_{k'},
\end{align}
where the first inequality comes from the triangle inequality,
the second inequality comes from the assumption,
and the last equality comes from the fact that
$\lrbar{\theta_{K+1}-1}\prod_{k'=K+1+1}^{K+1}\theta_{k'}=\lrbar{\theta_{K+1}-1}$.
Then the lemma is shown by induction.
\end{IEEEproof}

\begin{lem}[{\cite[Lemma 4]{CRNG-CHANNEL}\cite[Corollary 2]{SDECODING}}]
\label{lem:sdecoding}
Let $(U,V)$ be a pair consisting of state $U\in\U$ and observation $V\in\V$,
where $\mu_{UV}$ is the joint distribution of $(U,V)$.
We make a stochastic decision with $\mu_{U|V}$
that guesses state $U$ by $\hU\in\U$,
that is, the joint distribution of $(U,V,\hU)$ is given as
\begin{equation*}
 \mu_{UV\hU}(u,v,\hu)\equiv\mu_{UV}(u,v)\mu_{U|V}(\hu|v).
\end{equation*}
Then the decision error probability of this rule
is at most twice the decision error probability
of {\it any} (possibly stochastic) decision, 
that is,
\begin{align*}
 \sum_{\substack{
   u\in\U,v\in\V,\hu\in\U:
   \\
   \hu\neq u
 }}
 \mu_{UV}(u,v)
 \mu_{U|V}(\hu|v)
 &\leq
 2
 \sum_{\substack{
   u\in\U,v\in\V,\chu\in\U:
   \\
   \chu\neq u
 }}
 \mu_{UV}(u,v)
 \mu_{\chU|V}(\chu|v)
\end{align*}
for any arbitrary probability distribution $\mu_{\chU|V}$.
\end{lem}
\begin{IEEEproof}
 Here, we show the lemma directly for the completeness of this paper.
 We have
 \begin{align}
  \sum_{\substack{
    u\in\U,v\in\V,\hu\in\U:
    \\
    \hu\neq u
  }}
  \mu_{UV}(u,v)
  \mu_{U|V}(\hu|v)
  &=
  \sum_{\substack{
    u\in\U,v\in\V
  }}
  \mu_{UV}(u,v)
  [1-\mu_{U|V}(u|v)]
  \notag
  \\
  &=
  \sum_{\substack{
    u\in\U,v\in\V
  }}
  [\mu_{U|V}(u|v)-\mu_{U|V}(u|v)^2]
  \mu_V(v)
  \notag
  \\
  &\leq
  \sum_{\substack{
    u\in\U,v\in\V
  }}
  [\mu_{U|V}(u|v)-\mu_{U|V}(u|v)^2]
  \mu_V(v)
  +\sum_{\substack{
    u\in\U,v\in\V
  }}
  [\mu_{U|V}(u|v)-\mu_{\chU|V}(u|v)]
  \mu_V(v)
  \notag
  \\*
  &\quad
  +\sum_{\substack{
    u\in\U,v\in\V
  }}
  [\mu_{U|V}(u|v)-\mu_{\chU|V}(u|v)]^2
  \mu_V(v)
  +\sum_{\substack{
    u\in\U,v\in\V
  }}
  \mu_{\chU|V}(u|v)[1-\mu_{\chU|V}(u|v)]
  \mu_V(v)
  \notag
  \\
  &=
  \sum_{\substack{
    u\in\U,v\in\V
  }}
  2\mu_{U|V}(u|v)
  [1-\mu_{\chU|V}(u|v)]
  \mu_V(v)
  \notag
  \\
  &=
  2\sum_{\substack{
    u\in\U,v\in\V
  }}
  \mu_{UV}(u,v)
  [1-\mu_{\chU|V}(u|v)]
  \notag
  \\
  &=
  2\sum_{\substack{
   u\in\U,v\in\V,\chu\in\U:
   \\
   \chu\neq u
 }}
 \mu_{UV}(u,v)
 \mu_{\chU|V}(\chu|v),
\end{align}
where the inequality comes from the fact that
$\sum_{u\in\U}\mu_{U|V}(u|v)=\sum_{u\in\U}\mu_{\chU|V}(u|v)=1$,
and $\mu_{\chU|V}(u|v)\in[0,1]$ for all $u\in\U$ and $v\in\V$.
\end{IEEEproof}

\end{document}